\def\zem{$z_{\rm em}$}
\def\zabs{$z_{\rm abs}$}

\def\ha{H-$\alpha$}

\def\lya{Ly-$\alpha$}

\def\nii{[N~{\sc ii}]}

\def\hi{H~{\sc i}}
\def\hii{H~{\sc ii}}

\def\nhi{\mbox{$\sc N(\sc H~{\sc I})$}}
\def\lognhi{\mbox{$\log \sc N(\sc H~{\sc I})$}}

\def\nhi{\mbox{$\sc N(\sc H~{\sc I})$}}
\def\lognhi{\mbox{$\log \sc N(\sc H~{\sc I})$}}

\def\cii{C~{\sc ii}}
\def\ciii{C~{\sc iii}}

\def\feii{Fe~{\sc ii}}

\def\mgi{Mg~{\sc i}}
\def\mgii{Mg~{\sc ii}}

\def\nii{[N~{\sc II}]}
\def\naid{Na~{\sc ID}}
\def\sii{S~{\sc i}}

\def\aliii{Al~{\sc iii}}

\def\sii{S~{\sc ii}}

\documentclass[useAMS,epsfig]{mn2e}

\usepackage{rotating}

\usepackage{graphicx}

\title[Masses and Gas Flows ]{A SINFONI Integral Field Spectroscopy Survey for Galaxy Counterparts to Damped Lyman-$\alpha$ Systems - IV. Masses and Gas Flows\thanks{Based on observations collected during programme ESO 87.A-0414 at the European Southern Observatory with SINFONI and X-Shooter on the 8.2 m telescopes operated at the Paranal Observatory, Chile.} }
\author[C\'eline P\'eroux et al.] {C\'eline P\'eroux$^1$\thanks{e-mail:celine.peroux@gmail.com}, Nicolas Bouch\'e$^{2,3}$, Varsha P. Kulkarni$^4$ \& Donald G. York$^5$\\
$^1$ Aix Marseille Universit\'e, CNRS, LAM (Laboratoire d'Astrophysique de Marseille) UMR 7326, 13388, Marseille, France.  \\
$^2$ CNRS; Institut de Recherche en Astrophysique et Plan«etologie [IRAP] de Toulouse, 14 Avenue E. Belin, F-31400 Toulouse, France.\\
$^3$ Universit«e Paul Sabatier de Toulouse; UPS-OMP; IRAP; F-31400 Toulouse, France.\\
$^4$ Dept. of Physics and Astronomy, Univ. of South Carolina, Columbia, SC 29208, USA.\\
$^5$ Dept. of Astronomy and Astrophysics and The Enrico Fermi Institute, University of Chicago, 5640 S. Ellis Ave, Chicago, IL 60637, USA.\\
}
\begin{document}

\date{Accepted 2013 September 17.  Received 2013 September 16; in original form 2012 October 26}

\pagerange{\pageref{firstpage}--\pageref{lastpage}} \pubyear{2002}

\maketitle

\label{firstpage}

\begin{abstract}
The circumgalactic medium (CGM) of typical galaxies is crucial to our understanding of the cycling of gas into, through and out of
galaxies. One way to probe the CGM is to study gas around galaxies detected via the absorption lines they produce in the spectra of background quasars.  Here, we present medium resolution and new $\sim$0.4-arcsec resolution ($\sim$3 kpc at z$\sim$1) 3D observations with VLT/SINFONI of galaxies responsible for high-\nhi\ quasar absorbers. These data allow to determine in details the kinematics of the objects: the four  z$\sim$1 objects are found to be rotation-supported as expected from inclined discs, while the fifth z$\sim$2 system is dispersion-dominated. Two of the systems show sign of interactions and merging. In addition,  we use several indicators (star formation per unit area, a comparison of emission and absorption kinematics, arguments based on the inclination and the orientation of the absorber to the quasar line-of-sight and the distribution of metals) to determine the direction of the gas flows in and out of these galaxies. In some cases, our observations are consistent with the gas seen in absorption being due to material co-rotating with their halos. In the case of absorbing-galaxies towards Q1009$-$0026 and Q2222$-$0946, these indicators point toward the presence of an outflow traced in absorption. 
\end{abstract}

\section{Introduction}

The next challenge in galaxy evolution studies is to understand the physical processes involved in the formation of galaxies and their interactions with the medium surrounding them. The missing link is information about gas in the circumgalactic medium (CGM) which is believed to be both the repository of the inflowing gas and the receptacle
of energy and metals generated inside the galaxy. 
Understanding the nature and evolution of this material over cosmic time is a crucial element of galaxy formation theory, as it is the main source of fuel for star formation and is the material that is most sensitive to the ill-understood feedback processes that regulate galaxy growth. Of particular importance are the processes through which these galaxies accrete gas and subsequently form stars (Putman et al. 2009). 
The accretion of baryonic gas is complex and a range of models including infalling gas via cold mode accretion along intergalactic filaments (White \& Rees 1978; Dekel et al. 2009) can explain the observations. 
In addition, outflows are required to explain many observables including the mass-metallicity relation (hereafter MMR). This correlation indicates that galaxies with higher stellar masses have higher metallicity than less massive galaxies (Lequeux et al. 1979). The mass-metallicity relation is known to evolve with redshift (Tremonti et al. 2004; Savaglio et al. 2005; Erb et al. 2006; Maiolino et al. 2008) from z$\sim$0 to z $\sim$2--3. This relation has far-reaching consequences and is now an important ingredient of any galaxy evolution model (Kobayashi et al. 2007; Dav\'e \& Oppenheimer 2007). This correlation has also been studied in quasar absorbers. Indeed, several authors have reported an empirical relation between velocity spread and metallicity (P\'eroux et al. 2003; Ledoux et al. 2006; Meiring et al. 2007; Fynbo et al. 2008; M\"oller et al. 2013). However, the interpretation of these observations heavily relies on the assumption that the velocity dispersion is related to the mass of the absorbing galaxies (Haehnelt, Steinmetz \& Rauch 1998; Maller et al. 2001). Indeed, this interpretation is challenged by observations of \hi-gas rich local galaxies by Zwaan et al. (2008) which show that the velocity spread is
not a good indicator of mass in \hi-gas rich DLA analogues at z =0. Similarly, Bouch\'e et al. (2007) have suggested that the
equivalent width of \mgii\ absorbers is due to winds, not gravity
related velocity dispersion. Direct constraints on this relation in high-redshift quasar absorbers are only starting now with new measurements of stellar masses in absorbing-galaxies (P\'eroux et al. 2011b; Krogager et al. 2013; Fynbo et al. 2013). In addition, 3D observations (e.g., P\'eroux et al. 2011a, 2012) allow for direct studies of the dynamical and halo mass of absorbing-galaxies and comparison with expectations from models such as those by Pontzen et al. 2008. Besides the MMR, the presence of starburst-driven galactic winds is well established (e.g. Heckman 2003). The metal-pollution of the low-density intergalactic medium up to high-redshift (Pettini et al. 2003; Ryan-Weber et al. 2009) provides an additional argument in favour of the presence of winds in high-redshift galaxies.

Similarly, the accretion phenomenon is supported both by models and indirect observations. Indeed, the quantity of neutral \hi\ gas with cosmic time (Noterdaeme et al. 2012; Zafar et al. 2013) is found to be almost constant between z$\sim$5 to z$\sim$0.5, thus arguing for accretion to balance out the observed star formation history of the Universe (e.g. Cucciati et al. 2012). 

Given these expectations, direct observational probes of outflow and gas infall are required. Galactic winds are commonly traced via the blueshifts of interstellar absorption lines from cool gas traced by \naid\ and \mgii\ doublets super-imprinted on the stellar continuum (Shapley et al. 2003; Weiner et al. 2009; Steidel et al. 2010; Tremonti et al. 2007; Martin \& Bouch\'e 2009;
Rubin et al. 2010; Coil et al. 2011; Martin et al. 2012). Galactic-scale outflows are found to extend
along the galaxy minor axis (Bordoloi et al.
2011). Shapley et al. (2003) have shown that strong outflows are ubiquitous in galaxies of all morphological types at z$\sim$2--3 (see Law et al. 2012 for more recent results). Moreover, Bouch\'e et al. (2012) have shown that a bimodal distribution of the azimuthal orientation of the quasar sight lines with strong \mgii\ absorbers allows one to distinguish winds from gas associated with the disc among this population of absorbers (see also Kacprzak et al. 2012b). 
Interestingly, outflows have also been probed in emission (Steidel et al. 2011; Newman et al. 2012; Martin et al. 2013).

While observational evidences for outflow are growing at both low- and high-redshifts,  direct indicators of infall are notoriously more difficult to gather. Nevertheless, cool gas inflows have recently been detected in a few galaxies
(Sato et al. 2009; Giavalisco et al. 2011). Gas traced by \mgii\ has been observed to be infalling (Martin et al.
2012) with velocities of 100--200 km/s (Rubin et al. 2011) but in only 5\% of the galaxies. 
Recently, Bouch\'e et al. (2013) find strong evidence for cold accretion based on both a comparison of the kinematical properties and metallicities probed in emission and in absorption in a high-\nhi\ system.

A way to study the detailed processes at play in the CGM is to bring together, in a unified picture, data 
on cold gas ($<$100,000 K), metals and stellar content of the same galaxies. 
Indeed, the gas from the diffuse medium surrounding galaxies, is detected via
the absorption lines it produces in the spectra of background quasars, and
provides a powerful tool to study the CGM of galaxies (Stewart et al. 2011;  Stinson et al. 2012). 
Samples of the strongest of these quasar absorbers, 
the so-called Damped Lyman-$\alpha$ systems (DLAs), 
now amount to several hundreds (Prochaska et al. 2005; Noterdaeme et al. 2009; Noterdaeme et al. 2012; York et al. in prep) and the number of known sub-Damped Lyman-$\alpha$ systems (sub-DLAs; P\'eroux et al. 2003) is also growing (Zafar et al. 2013). These \hi-selected sight-lines offer the prospect to study the direct surroundings of intermediate-redshift galaxies in the few cases where the absorbing-galaxy has been identified.

With the aim of studying the flows of gas in and out of galaxies, Bouch\'e et al. (2007a) have taken advantage of the 3D spectroscopy at near-IR wavelengths made possible by  SINFONI on VLT to successfully detect the galaxies responsible for \mgii\ absorbers. 
In addition, P\'eroux et al. (2011a; 2012) have been able to detect five high-N(\hi) absorbing-galaxies out of 16 searched for at $z \simeq 1$--2. In these studies,using SINFONI data at a resolution of 0.8-arcsec ($\sim$6 kpc at z$\sim$1), we identified galaxy counterparts for the absorbers and estimated star formation rates and emission metallicities from emission line detections. Here, we present new SINFONI data, at a resolution of 0.4-arcsec ($\sim$3 kpc at z$\sim$1), of four of the five detected sub-/DLAs in the sample. These data, in combination our earlier observations at a resolution of 0.8-arcsec, allow us to estimate the masses of the systems, to study the spatially resolved kinematics and therefore the dynamical state of these galaxies.

The present paper is structured as follows. A summary of observational details and data reduction steps are provided in Section 2. 
In the third section, we detail the kinematical analysis for each absorbing galaxy including mass estimates. We present several lines of evidence which help to characterize the flow of gas around these galaxies in Section 4. Throughout this paper, we assume a cosmology with H$_0$=71 km/s/Mpc, $\Omega_M$=0.27 and $\Omega_{\rm \lambda}$=0.73.

\section{Observations and Data Reduction}

\begin{table*}
\begin{center}
\caption{{\bf Journal of high-resolution SINFONI observations.} The data are a 3 arcsec $\times$ 3 arcsec field of view, corresponding to a 0.1-arcsec pixel-scale.}
\label{t:JoO}
\begin{tabular}{cccccccccc}
\hline\hline
Quasar 		  &Coordinates$^a$ &V Mag &\zem &\zabs  &Observing Date &T$_{\rm exp}$[sec]$\times $N$_{\rm exp}$$^b$ &Band &AO$^c$ &PSF ["] \\
\hline
Q0302$-$223 		&03 04 49.86 $-$22 11 51.9	&16.0            &1.409	&1.0094           &2011 Sep 12/27 &900$\times$(4+2) &J  &NGS		&0.35	\\	
Q0452$-$1640 	&04 52 13.60 $-$16 40 12.0	&18.0     	    &2.679	&1.0072         &2011 Sep  22 &600$\times$8	        &J  &no AO	&0.40	\\	
Q2222$-$0946 	&SDSSJ222256.11$-$094636.2 	&18.3            &2.927	&2.3543         &2011 Aug 6/10/28+Sep   &900$\times$(20+20) &K &NGS	&0.40\\
... 	&... 	&...            &... 	&...     & 7/26/29/30+Oct 1/2   &+600$\times$(8+4) &... &...	&...	\\	
Q2352$-$0028 	&SDSSJ235253.51$-$002850.4 	&18.6            &1.624	&1.0318         &2011 Aug 6/22+Sep 29  &900$\times$(20+12)&J  &NGS		&0.40	\\	
\hline\hline 				       			 	 
\end{tabular}			       			 	 
\end{center}			       			 	 
\begin{minipage}{180mm}
{\bf Note:} \\
{\bf $^a$} SIMBAD coordinates unless the quasars is part of SDSS, in which case SDSS names are provided.\\
{\bf $^b$} the two numbers for N$_{\rm exp}$ in brackets refer to exposures classified as 'completed' and 'executed' (i.e. not within the user specifications in ESO terminology), respectively.\\  
{\bf $^c$} no AO: no Adaptive Optics, natural seeing. NGS: Adaptive Optics with a Natural Guide Star.\\
\end{minipage}
\end{table*}			       			 	 

We have reobserved four of the five \nhi-systems discovered with SINFONI in P\'eroux et al. (2011a, 2012) with dedicated observations centered on the absorbing-galaxy with known sky position and with higher spatial resolution. Observing time was awarded for the fifth detected system (Q1009$-$0026) as well, but those observations could not be obtained. A journal of observations summarising the target properties and experimental set-up is presented in Table~\ref{t:JoO}. The table provides the observing date and exposure times for each of the objects and the resulting PSF of the combined data. The observations were carried out in service mode (under programme ESO 87.A-0414) at the European Southern Observatory with SINFONI on the 8.2 m YEPUN telescope. The redshifted \ha\ line lies in the $J$-band for the three targets at z$\sim$1 and in the $K$-band for the fourth system at z$\sim$2 (Q2222$-$0946). For fields where the quasar itself is bright enough or in which a bright star is available nearby, we have used it as a natural guide star (NGS) for adaptive optics (AO) in order to improve the spatial resolution. For the field towards Q0452$-$1640 with no suitable tip-tilt stars, we were awarded laser guide star (LGS) AO observations but the data were taken with natural seeing (no AO). The resulting PSF measured on the quasars in the cubes ranges from 0.35" to 0.40", considerably better than our 0.25-arcsec pixel-scale data published previously (for which the PSF ranged from 0.6 to 1.1 arcsec). The $J$ ($K$) grism provides a spectral resolution of around R$\sim$2000 (4000). The resulting cubes (a mosaic of the 3"$\times$3" SINFONI field-of-view resulting in a 5"$\times$5" effective field-of-view) are centered on the absorbing-galaxy with known sky positions.

The data were reduced with the latest version of the ESO SINFONI pipeline (version 2.3.2) and custom routines. The latter were used to correct the raw cubes for detector bad columns and to remove cosmic rays by applying the Laplacian edge technique of van Dokkum (2001). Master bias and flat images based on calibration cubes taken closest in
time to the science frames were used to correct each data cube. Bias and flat-field correction were done within the ESO pipeline with additional OH line suppression and sky subtraction using additional codes. Within one Observing Block, the science frames were pair-subtracted with an ON-OFF pattern to eliminate variation in the infra-red sky background. The wavelength calibration was based on the Ar lamp and is accurate to about $\sim$ 30 km/s in the J-band, i.e. comparable with the calculated heliocentric correction (of the order of 10-30 km/s). For each set of observations, a flux standard star was observed at approximately the same time, with similar airmass and was reduced in the same way as the science data. These standard stars are then used for flux calibration by fitting a black body spectrum to the O/B stars or a power law to the cool stars (T$<$10,000K) and normalising them to the 2MASS magnitudes. These spectra were also used to remove atmospheric absorption features from the science cubes. When a quasar was included in the cube, the resulting flux calibration is compared with the quasar 2MASS magnitudes in order to estimate the flux uncertainties. The different observations from the independent Observing Blocks were then combined spatially using the position of the quasar in each frame resulting in an average co-added cube per target. In one case (Q2352$-$0028), there has been an offset in the centering of the field at the telescope and the quasar is not covered by the observations, thus complicating this step.

\section{Kinematics and Mass Estimates}

\subsection{Methodology}

For each target, we extracted maps of the velocity-integrated line fluxes, relative velocities, and velocity dispersion from the SINFONI data cubes. The maps were built using standard procedures. Namely, for each pixel, we fitted the line profile to a Gaussian convolved with a template of the instrument profile following e.g., F\"orster-Schreiber et al. (2006, 2009). Thus, the fitting takes into account the instrumental line profile. In addition, weighted fits are performed using Monte Carlo resampling based on the spectral noise profile. 

In order to perform the dynamical study of the systems, we apply the new 3D parametric model developed by Bouch\'e et al. (2013). This technique presents an improvement over the more traditional 2D modelling because it simultaneously optimizes 9 parameters without prior assumptions and because it uses all of the information available (10,000 pixels) most of which have very low signal-to-noise. Our code fits a 3D (x,y,lambda) model directly to the data using a Baysian technique. The model fits an exponential flux profile to the observed data and either an arctangent or an exponential law to the mass profile i.e. assuming circular orbits. The fitted parameters include the two spatial and one spectral dimensions of the dynamical centre, the scale-length of the exponential profile fitted to the observed flux distribution, the total flux in the line, the inclination of the galaxy, its position angle, maximum circular velocity and intrinsic dispersion of the object. Note, all the
values derived are intrinsic (i.e. should not be corrected for the inclination) since they refer to the rotating disk in 3Dxyz space before projecting the velocities along the line-of-sight. The inclination of the galaxy is setting the axis ratio of the model. The routine is based on a Monte-Carlo Markov-Chain (MCMC) optimization run over 5000 iterations. Further tests of this model, including a comparison with results from 2D analyses and additional details can be found in Bouch\'e et al. (2013). The results of our analysis are summarised in the following sections for each of the individual systems.

\subsection{Q0302$-$223, \zabs=1.0094}

The DLA-galaxy, associated with the absorber with column density \lognhi=20.36$\pm$0.11, was observed in the early work of P\'eroux et al. (2011a). Additional archival HST/WFPC2 images covering this field (Le Brun et al. 1997) indicated that the system is composed of two individual components. The two separate components are not resolved in the low-resolution SINFONI observations of P\'eroux et al. (2011a) but are consistent with the detected elongated shape. More surprisingly, the individual components are still not seen in the  new higher-resolution SINFONI data presented here. Indeed, we note that while the elongated shape is still apparent, the area over which the object is detected is not larger with the 0.1-arcsec pixel-scale than with the 0.25-arcsec pixel-scale observations. This is most probably due to the limits of detection in terms of surface brightness and illustrates the trade-off to be made between the spatial resolution and signal detection of intermediate redshift galaxies with IFU instruments. 

As a result, we refer to the kinematics of this system already presented in P\'eroux et al. (2011b). In summary, two possible explanations have been put forward to explain the data: one is that the observed dispersion is due to a small difference in redshift of the two interacting galaxies. We note however that this would not be consistent with the observed \ha\ light profile which appears to be exponential with a scale length of 0.6". So that the most probable explanation would be a dispersion-dominated galaxy with a disky morphology. Using the virial theorem, we deduce M$_{\rm dyn}$=10$^{10.3}$ $M_{\odot}$. The mass of the gas is found to be M$_{\rm gas}$=10$^{9.1}$ $M_{\odot}$ based on an inverted 'Schmidt-Kennicutt' law (Kennicutt et al. 1998). Given that this object shows evidence of a disc morphology without spiral arms feature, it could be a young S0 galaxy. All these values are tabulated in Table~\ref{t:kine}, which summarises the dynamical properties of all the absorbers in the sample.

\subsection{Q0452$-$1640, \zabs=1.0072}

\begin{figure*}
\begin{center}
\includegraphics[height=4.5cm, width=5.5cm, angle=0]{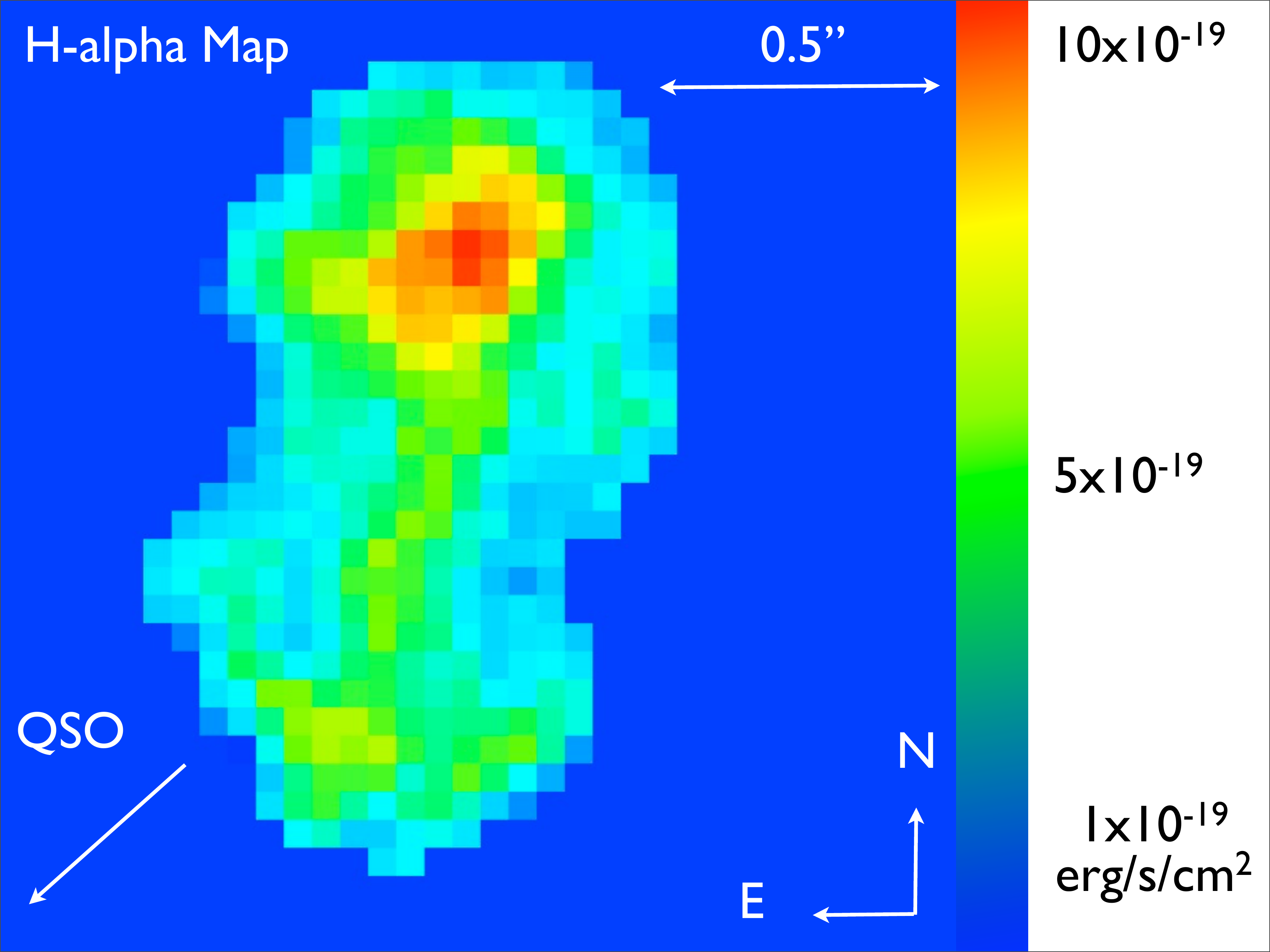}
\includegraphics[height=4.5cm, width=5.5cm, angle=0]{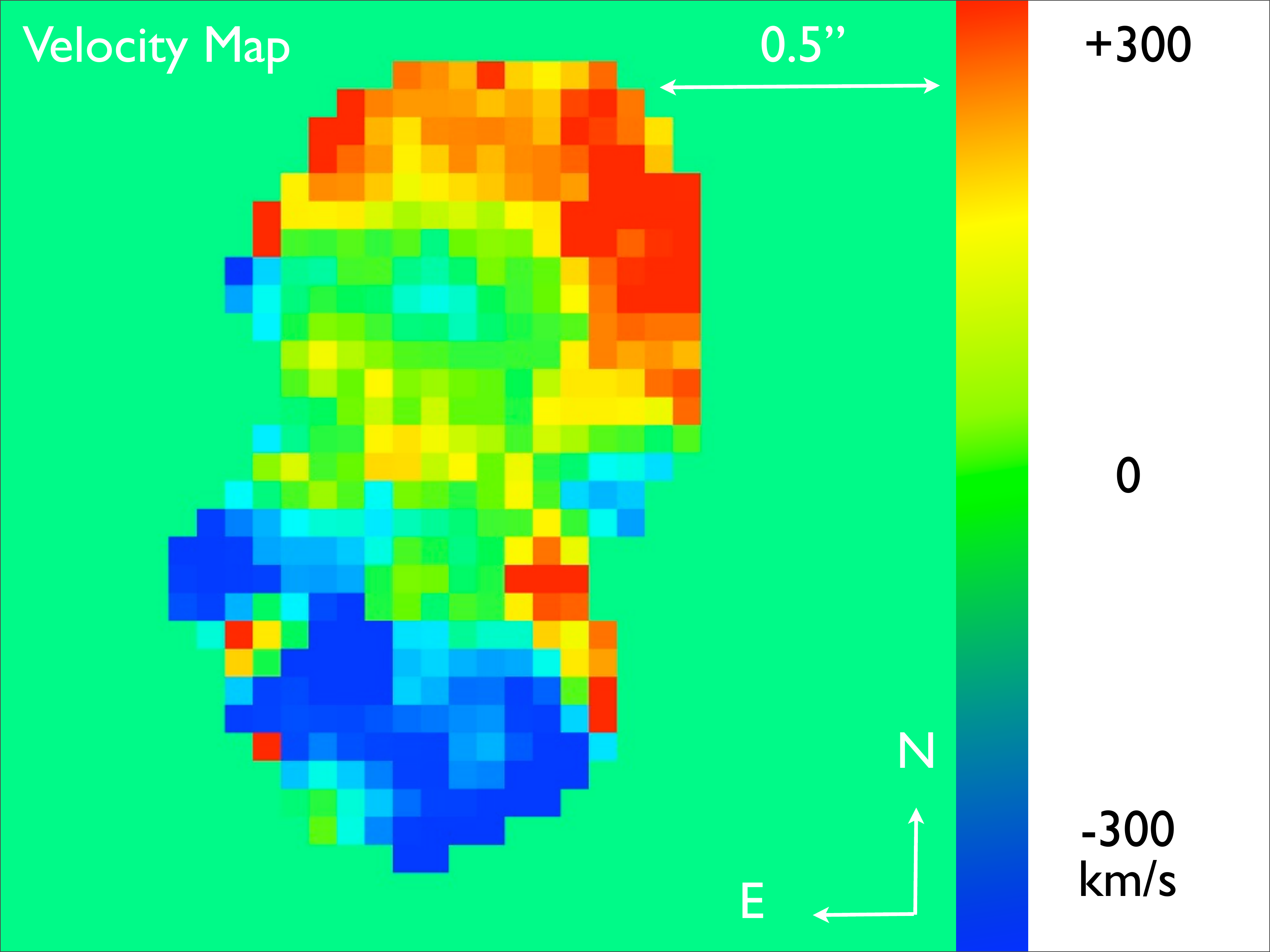}
\includegraphics[height=4.5cm, width=5.5cm, angle=0]{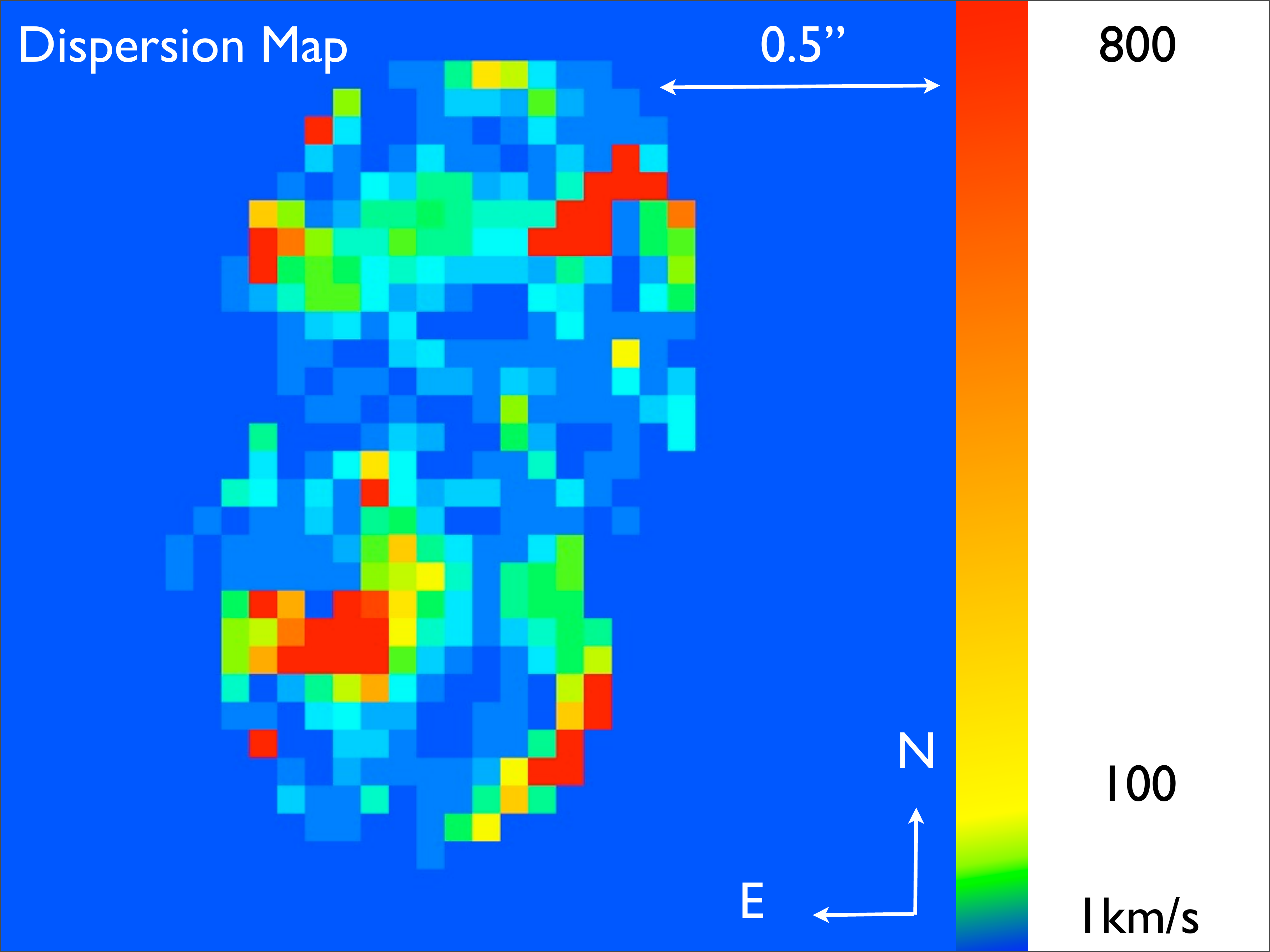}
\caption{{\bf \ha\ flux map, \ha\ velocity field and \ha\ velocity dispersion maps of the DLA-galaxy towards Q0452$-$1640 at  \zabs=1.0072.} In this and the following, the orientation and scales are indicated on the figures as well as the direction to the quasar. The new high-resolution SINFONI observations indicate that two objects are detected at the redshift of the DLA towards the quasar Q0452$-$1640. The system at the top, which is more extended, is also brighter. Its velocity map indicates a rotating object but the global dispersion is small, leading to an estimated v/$\sigma \gg 1$, typical of a normal disc. The bottom object shows an arc connected to the top galaxy by a filament emitting in \ha. This system resembles the ring galaxies recently reported by Genzel et al. (2008) which are thick discs with little instabilities in the centre. } 
\label{f:Q0452_kine}
\end{center}
\end{figure*}

\begin{figure*}
\begin{center}
\includegraphics[height=10cm, width=15cm, angle=0]{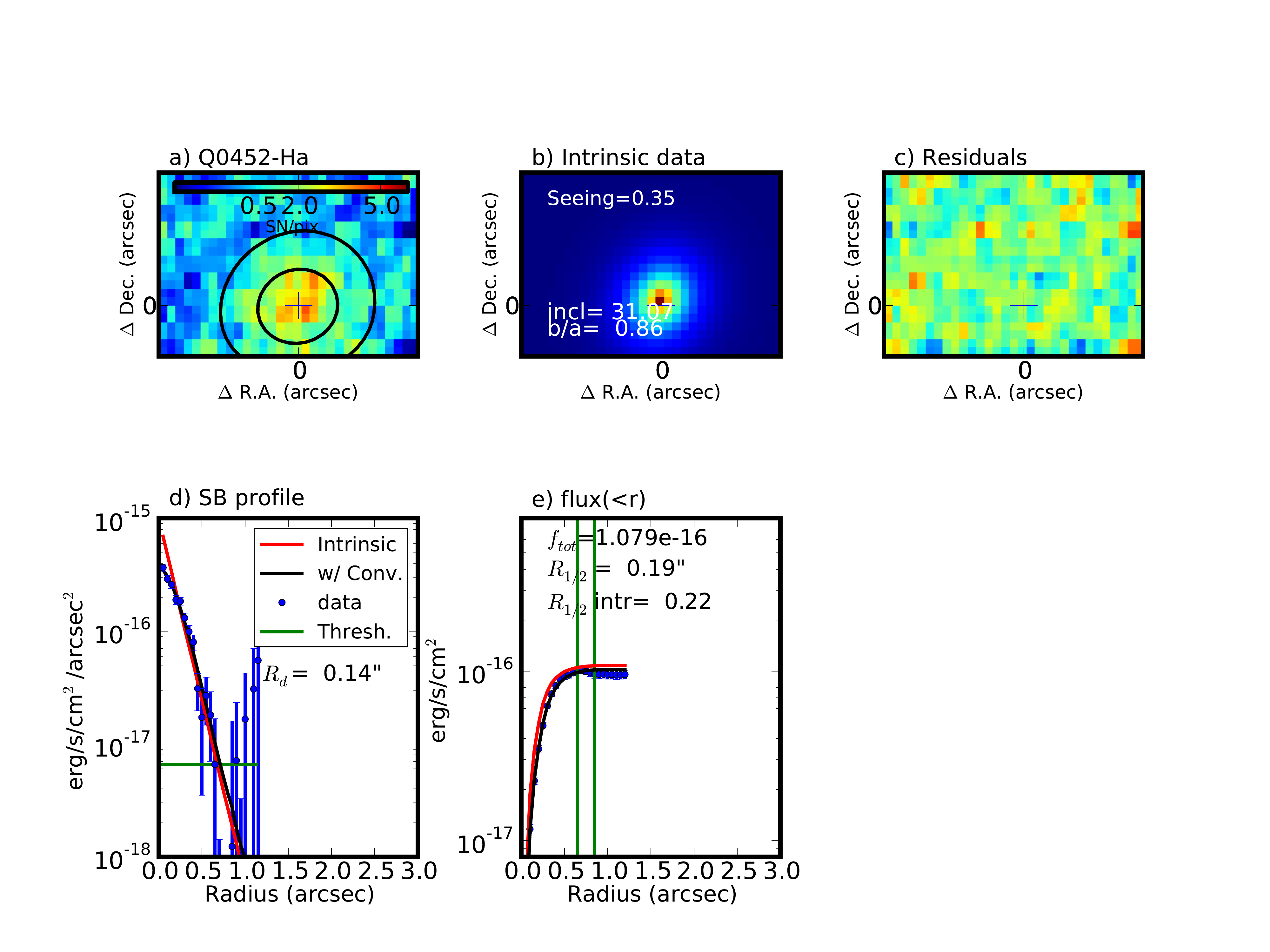}
\caption{{\bf Results from the fit to the 2D flux map.} In this figure as well as Figure~\ref{f:Q2222_3Dfit} and \ref{f:Q2352_3Dfit}, the top-left panel shows the observed \ha\ map colour-coded according to the SNR of the data and the top-middle panel presents the fitted 2D flux profile. To the right of that panel, the residuals between the data and the intrinsic profile convolved with the seeing are shown. The bottom-left panel presents the surface brightness profile summed over annuli of increasing radius. The data are shown with blue error bars while the red and black lines indicate the intrinsic and convolved fits. The observations are well modeled by the fit to the detection threshold. Finally, the bottom-middle panel presents the curve of growth of the flux in erg/s/cm$^2$ (f$_{<r}$) as a function of increasing radius. The green vertical lines indicate the region over which the total flux is measured. In the case of the absorber towards Q0452$-$1640 at  \zabs=1.0072, two distinct objects are detected. The dynamical analysis only converges for the system at the top which is presented here. The residual map for that fit shows no particular features. We derive sin i=0.50, corresponding to an inclination of 30 degrees. Because the galaxy is found to be almost face-on, the errors on the kinematic values estimates are large.}
\label{f:Q0452_3Dfit}
\end{center}
\end{figure*}

The DLA towards this quasar has an \hi\ column density \lognhi=20.98$^{+0.06}_{-0.07}$. The absorbing-galaxy was first reported by P\'eroux et al. (2012). From that low-resolution SINFONI data, an elongated shape in the direction of the quasar (to the south) was already apparent. The new high-resolution observations both confirm these findings and indicate the presence of two separate components. The kinematical analysis indicates that galaxy to the North is rotating, while the bottom object resembles a ring galaxy, albeit also with the indications of rotation. The full structure spans 1.5 arcsec, extended in the direction of the quasar. 

Figure~\ref{f:Q0452_kine} presents the \ha\ flux map, \ha\ velocity field and \ha\ velocity dispersion maps for the two objects detected towards Q0452$-$1640. The object at the top, which is more extended, is also brighter. The velocity map indicates a rotating object but the global dispersion is small, typical of a normal disc. The bottom object shows an arc connected to the top galaxy by a filament emitting in \ha. This system resembles the ring galaxies recently reported by Genzel et al. (2008) which are due to thick discs with little instabilities in the centre. The velocity map also indicates a rotating object. The continuum of that object is detected in our data cube. The \hi\ gas seen in absorption towards this quasar line-of-sight is therefore linked to the interaction between the two galaxies.

The modelling was applied to the object at the bottom, but the process did not converge because of the unusual shape of the object. We thus concentrated on the 3D fit of the object at the top, trimming the region of interactions between the two objects to reduce to a minimum the influence of the object at the bottom. Figure~\ref{f:Q0452_3Dfit} illustrates the results of the fit to the 2D flux map. The top-left panel shows the observed \ha\ map colour-coded according to the SNR of the data and the top-middle panel presents the fitted 2D flux profile. To the right of that panel, the residuals between the data and the intrinsic profile convolved with the seeing are shown. In the case of the absorbing-galaxy towards Q0452$-$164, the residual map shows no particular features, indicating that the fit is robust. The bottom-left panel presents the surface brightness profile summed over annuli of increasing radius. The data are shown with blue error bars while the red and black lines indicate the intrinsic and convolved fits. The observations are well modeled by the fit until the detection threshold. Finally, the bottom-right panel presents the curve of growth of the flux in erg/s/cm$^2$ (f$_{<r}$) as a function of increasing radius. The green vertical lines indicate the region over which the total flux is measured and the results are in good agreement with estimates based on SINFONI 1D spectra made by P\'eroux et al. (2012). We derive sin i=0.50 from the ratio of minor to major axes, corresponding to an inclination of 30 degrees.

Because the galaxy is found to be almost face-on, the errors on the kinematic value from the 3D analysis are large. In particular, the maximum circular velocity measured between the minimum and maximum values of the velocity field is poorly constrained. We derive 2 $V_{\rm max}$ sin i$\sim$100km/s, corresponding to an inclination-corrected maximum velocity of $V_{\rm max} \sim$100
km/s. The global dispersion is large and typical of a normal disc leading to an estimated v/$\sigma \gg 1$. Finally, 
we derive a half-light radius r$_{1/2}=$0.2" ($\sim$2kpc). Since no continuum is detected in our data, the half-light radii provided are measured on the \ha\ light. Nelson et al. (2012) have shown that \ha\ sizes generally track the stellar continuum sizes but are typically a factor of 1.3 larger. This would produce a systematic offset in the calculated M$_{\rm dyn}$ with respect to values from the literature.

Since we know that the system is rotating, we can use the enclosed mass to determine the dynamical mass (following e.g. Epinat et al. 2009):

\begin{equation}
M_{\rm dyn} = V_{\rm max}^2~r_{1/2}~/~G
\end{equation}

where $V_{\rm max}$ is the maximum velocity and r$_{1/2}$ is as before. We therefore derive M$_{\rm dyn}$=10$^{10.6}$ $M_{\odot}$. 
 In order to compute the gas mass, we start from the observed \ha\ surface brightness and compute gas surface brightness using an inverse 'Schmidt-Kennicutt' relation (Bouch\'e et al. 2007b; Finkelstein et al. 2009; Daddi et al. 2010; Genzel et al. 2010, 2011):

\begin{equation}
\Sigma_{\rm gas} [M_{\odot}/pc^{2}] = 1.6 \times 10^{-27}\left(\frac{\Sigma_{H\alpha}}{[erg/s/kpc^{2}]}\right)^{0.71} 
\end{equation}

In this case, we find a gas mass M$_{\rm gas}$=10$^{9.2}$ $M_{\odot}$, which is based on the mean of the gas surface density measured over the visible region. This, together with the 0.3 dex scatter in the Schmidt-Kennicutt relation, means that large uncertainties are associated with such measurement. We also note that the gas mass is significantly lower than the dynamical mass in this system, indicating a low gas fraction in the object.

Finally, we are able to estimate the mass of the halo in which the system resides, assuming a spherical virialised collapse model (Mo \& White 2002):

\begin{equation}
M_{\rm halo}= 0.1 H_o^{-1} G^{-1} \Omega_m^{-0.5} (1+z)^{-1.5} V_{\rm max}^3
\end{equation}

We derive M$_{\rm halo}$=10$^{12.8}$ $M_{\odot}$. All these values are listed in Table~\ref{t:kine}.

\begin{table*}
\begin{center}
\caption{{\bf Kinematic properties and mass estimates of all the \nhi\ absorbers detected with SINFONI. } Details of the calculations are provided in the text.}
\label{t:kine}
\begin{tabular}{ccccccccccccc}
\hline\hline
Quasar 		  &sin $i$ &v/$\sigma$   &r$_{1/2}$ & $\Sigma_{SFR}$  &$M_{\rm dyn}$      &$\Sigma_{\rm gas}$	&$M_{\rm gas}$	& $M_{\rm halo}$& $M_{*}$ &Morphology &Rotating? &Ref \\
		           &          &	&["]&[M$_{\odot}$/yr/kpc$^2$] &[M$_{\odot}$] 	&[M$_{\odot}$/pc$^2$]       &[M$_{\odot}$]        &[M$_{\odot}$] &[M$_{\odot}$] & & &\\
\hline
Q0302$-$223{\bf $^*$}	&0.88	&0.19	&0.7 &0.13	&10$^{10.3}$	&10$^{1.9}$	&10$^{9.1}$	&--	&10$^{9.5}$	&merger	&yes & (a) \\
Q0452$-$1640{\bf $^\dagger$}  	&0.51	&$\gg$1	&0.2	&0.49	&10$^{10.6}$	&10$^{2.3}$	&10$^{9.2}$	&10$^{12.8}$&--	&merger/disc	&yes &(b)  \\
Q1009$-$0026	     		&0.60	&1.45	&0.5 &0.31	&10$^{10.9}$	&10$^{2.2}$	&10$^{9.2}$	&10$^{12.6}$&--   	&disc	&yes &(a)  \\
Q2222$-$0946  		&0.68	&0.35	&0.2	&1.97	&10$^{9.8}$	&10$^{2.8}$	&10$^{9.7}$	&-- 		&10$^{9.3}$	 	&compact	&no  &(b,c)  \\
Q2352$-$0028  		&0.93	&1.17	&0.7	&0.24	&10$^{10.4}$	&10$^{2.1}$	&10$^{8.8}$	&10$^{11.8}$&--	&disc	&yes &(b)  \\
\hline
\hline
\end{tabular}			       			 	 
\end{center}			       			 	 
\vspace{0.2cm}
\begin{minipage}{180mm}
{\bf References:} (a) P\'eroux et al. 2011b (b) This work (c) Krogager et al. (2013).\\
{\bf $^*$:} The HST/WFPC2 data from Le Brun et al. (1997) clearly show that the object is subdivided into two components, but even with the new higher-resolution 0.1-arcsec pixel-scale SINFONI observations presented here, these components are not resolved. Therefore, in this table, the object is treated as only one. \\
{\bf $^\dagger$:} The new higher-resolution 0.1-arcsec pixel-scale SINFONI data show that the object is subdivided into two components. The fit presented in this table refers to the rotation-dominated object to the North. \\
\end{minipage}
\end{table*}

\subsection{Q2222$-$0946, \zabs=2.3543}

\begin{figure*}
\begin{center}
\includegraphics[height=4.5cm, width=5.5cm, angle=0]{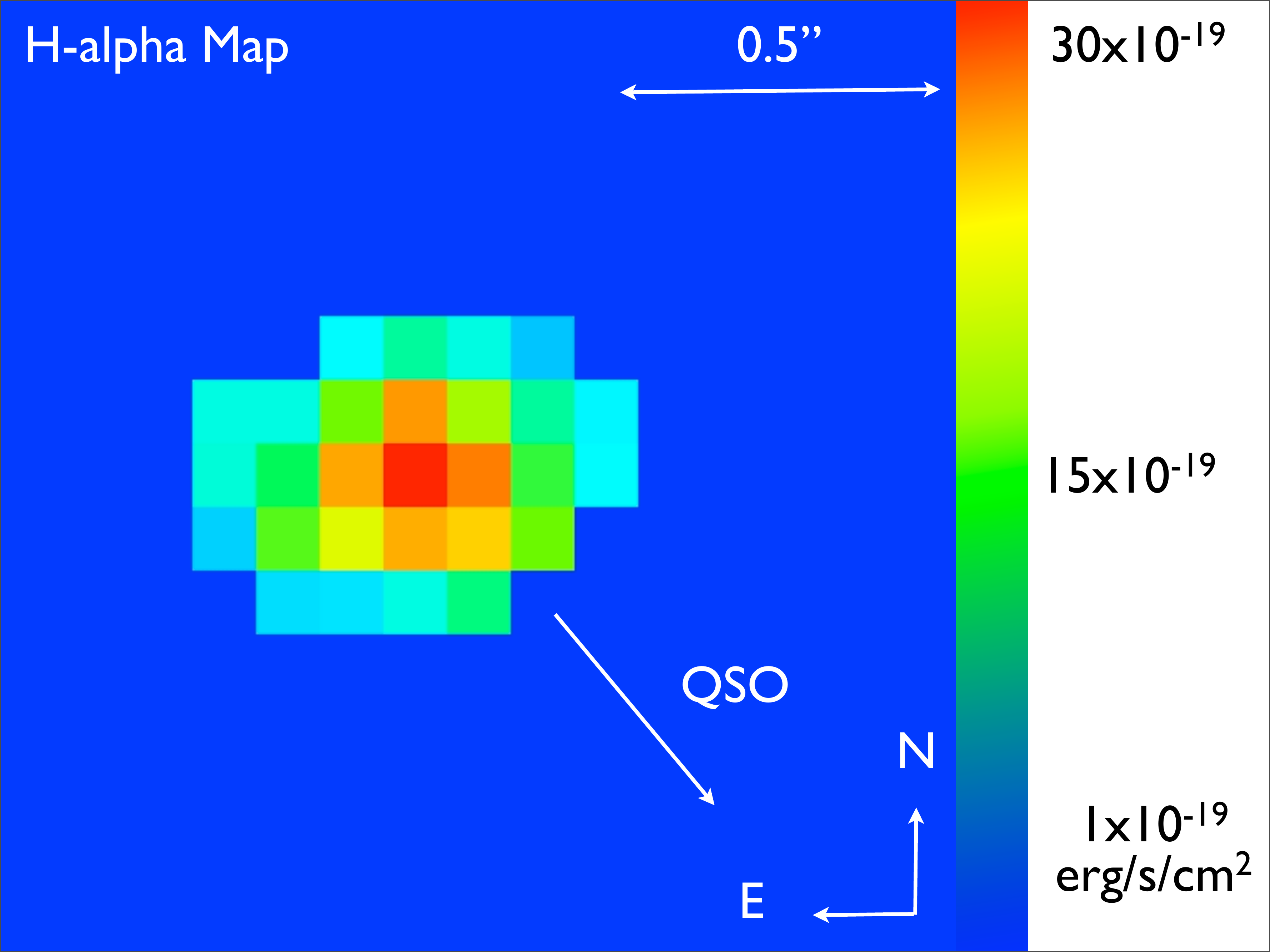}
\includegraphics[height=4.5cm, width=5.5cm, angle=0]{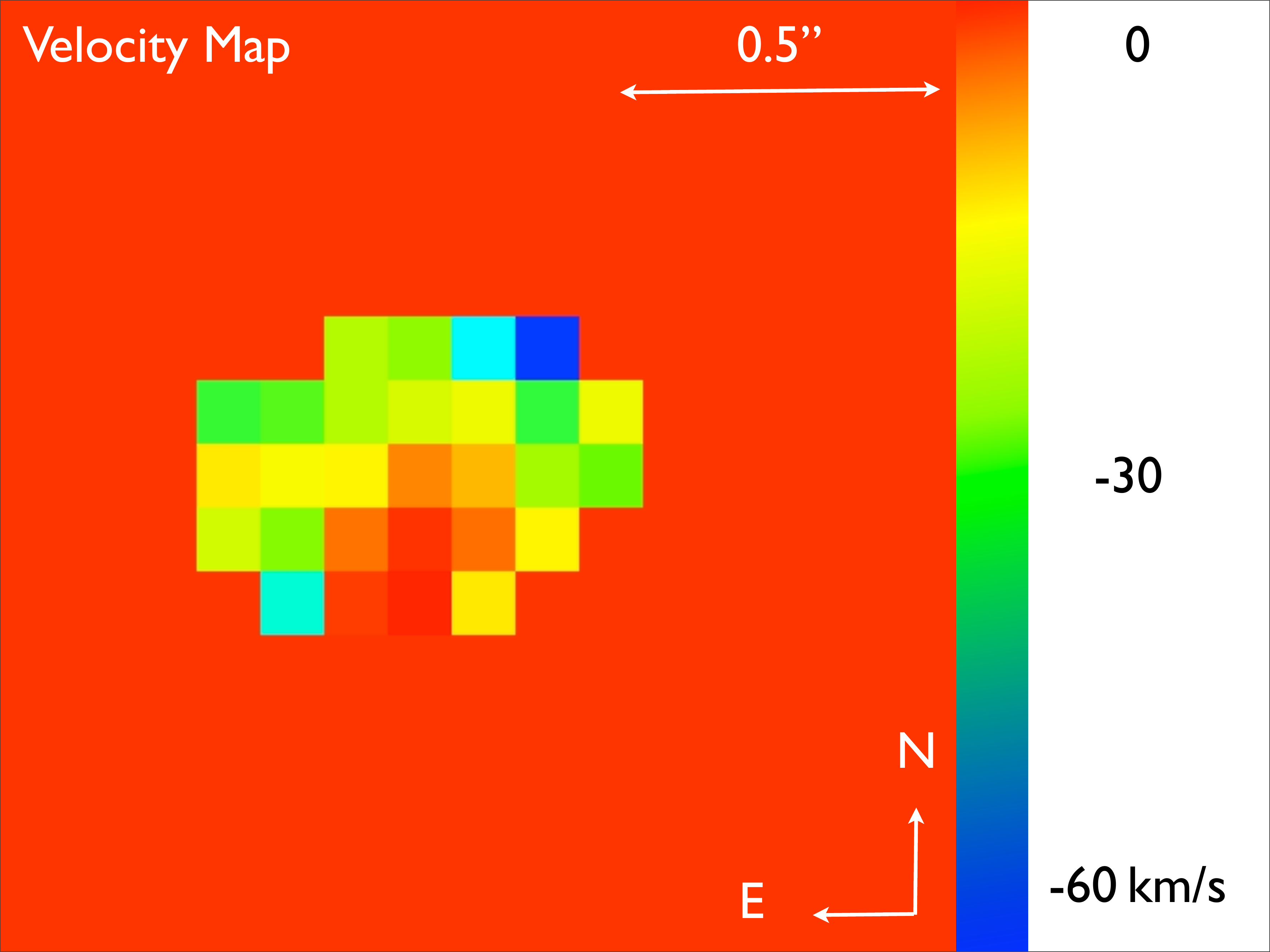}
\includegraphics[height=4.5cm, width=5.5cm, angle=0]{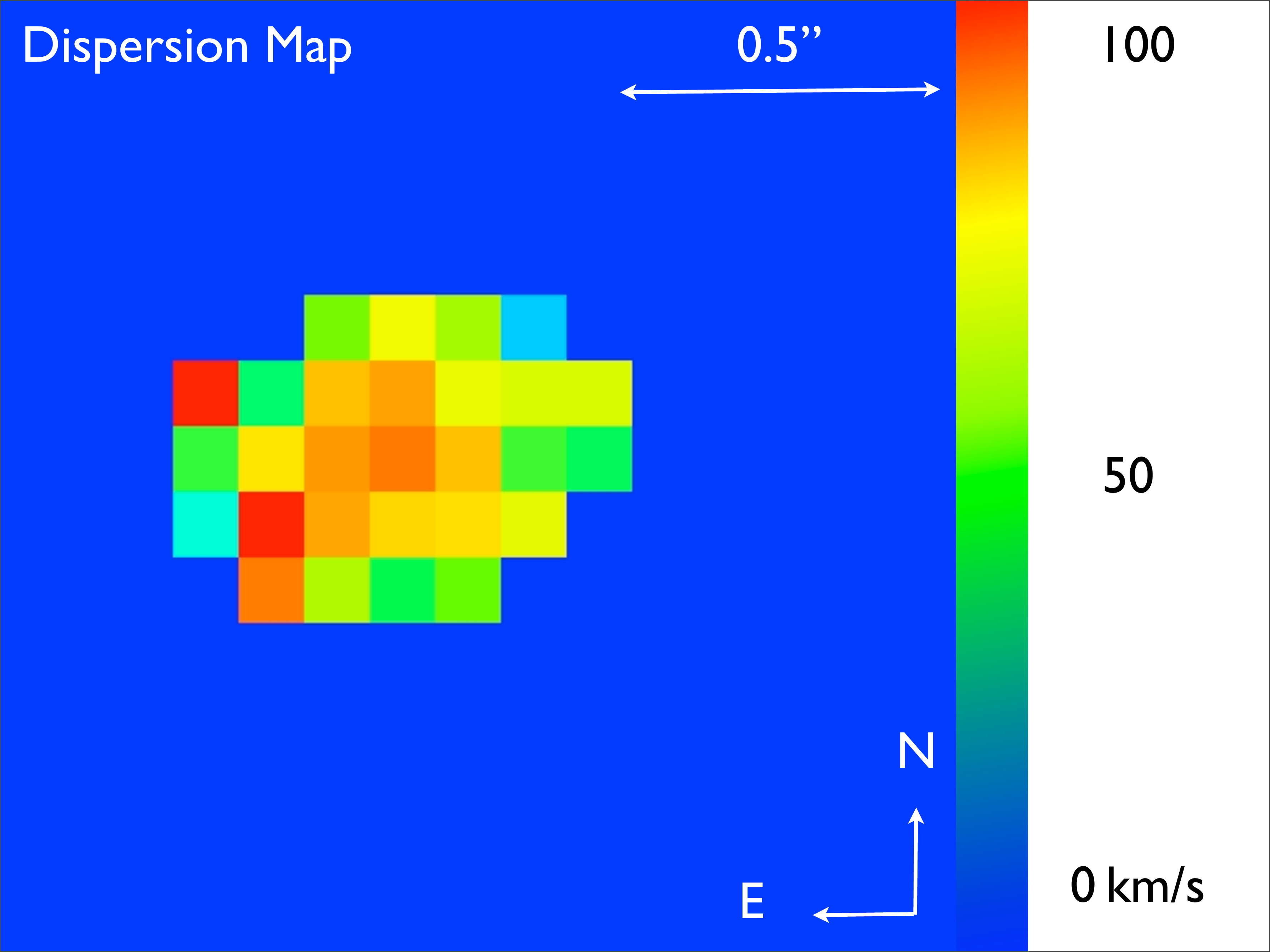}
\caption{{\bf \ha\ flux map, \ha\ velocity field and \ha\ velocity dispersion maps of the DLA-galaxy towards Q2222$-$0946  at \zabs=2.3543.} The velocity map clearly shows the absence of gradient in this case, but rather suggests a dispersion-dominated system. In fact, the dispersion at the centre of the object is found to be dominated by the PSF.}
\label{f:Q2222_kine}
\end{center}
\end{figure*}

\begin{figure*}
\begin{center}
\includegraphics[height=13cm, width=15cm, angle=0]{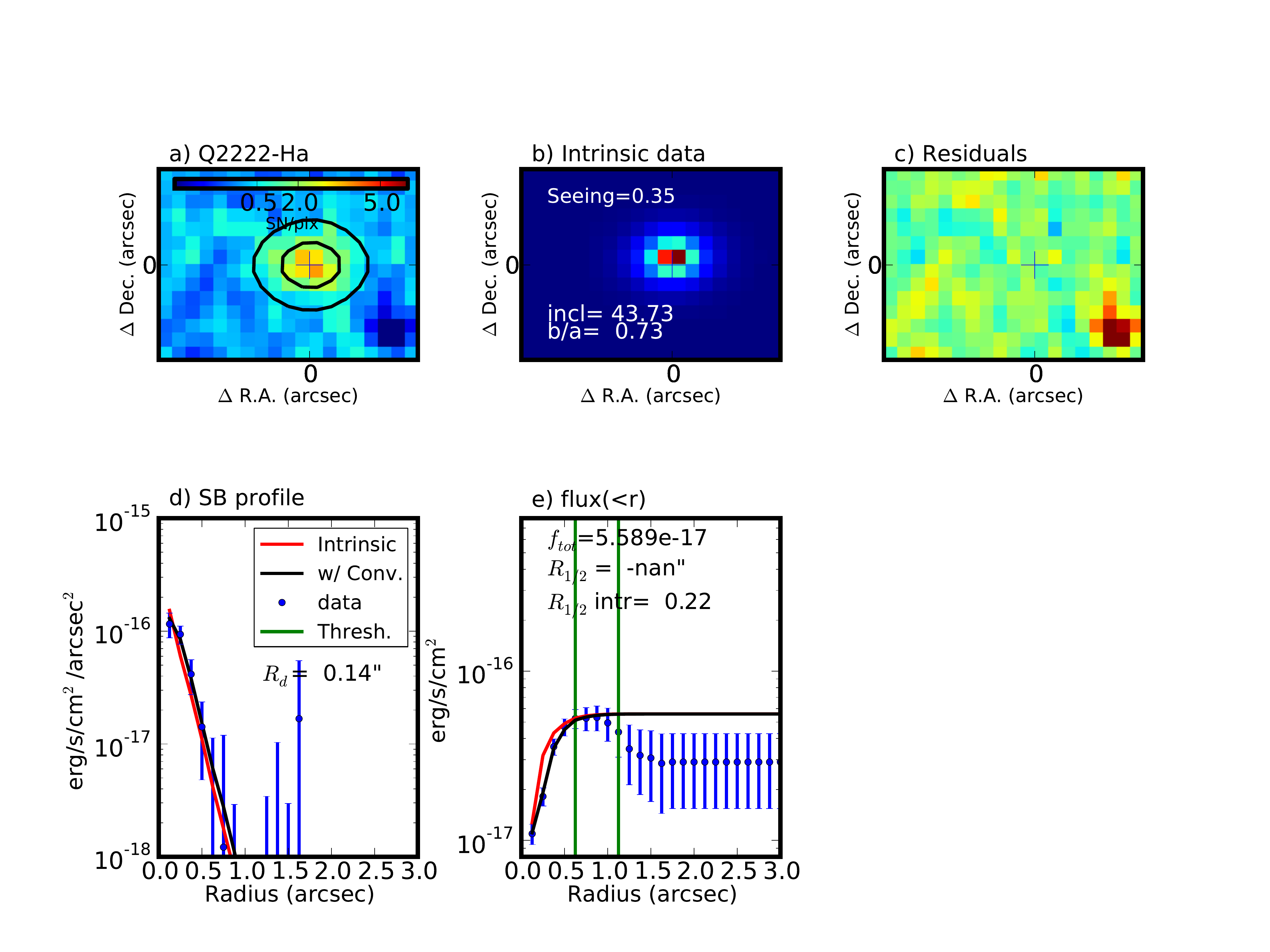}
\caption{{\bf Results from the fit to the 2D flux map of the absorber in the field of Q2222$-$0946 at \zabs=2.3543}. The fit for this system is complicated by the proximity on the sky of the bright background quasar. To avoid its contamination, we perform the analysis on the larger scale data and trim the cube at maximum. The residual from the quasar signature is visible in the residual map in the top-right panel of the figure. Because of the nature of the object, the kinematics are not well constrained in this case. }
\label{f:Q2222_3Dfit}
\end{center}
\end{figure*}

The DLA towards this quasar has an \hi\ column density \lognhi=20.50$\pm$0.15. The absorbing-galaxy was reported by Fynbo et al. (2010) from X-Shooter observations while P\'eroux et al. (2012) presented IFU observations of the object. From that low-resolution SINFONI data, the absorbing-galaxy is found to be very compact and close to the quasar (angular separation of 0.70" corresponding to a physical distance of 6 kpc at the redshift of the galaxy). The new high-resolution data confirm the compact morphology of the object and, as in the case for the absorbing-galaxy towards Q0302$-$223, does not extend further in radius than the lower-resolution data because of the limitation in surface brightness detection. Krogager et al. (2013) have recently detected the continuum of the object in Hubble Space Telescope F606W wide images and found it to be elongated along an East-West axis. This is not seen in the 0.1-arcsec pixel-scale SINFONI observations probably because the data are not sensitive enough to low surface-brightness signal with the given seeing (PSF = 0.40 arcsec).

Figure~\ref{f:Q2222_kine} presents the \ha\ flux map, \ha\ velocity field and \ha\ velocity dispersion maps in the 0.25-arcsec pixel-scale of the absorbing-galaxy detected towards Q2222$-$0946. The velocity map clearly shows the absence of a gradient in that case, but rather suggests a dispersion-dominated system. In fact, the dispersion at the centre of the object is found to be dominated by the seeing. The 3D kinematical fit for this system is complicated by the proximity of the bright background quasar. To avoid its contamination, we perform the analysis on the larger scale data and trim the cube at maximum. To avoid its contamination, we perform the analysis on the larger scale data and trim the cube at maximum. Figure~\ref{f:Q2222_3Dfit} illustrates the results of the fit to the 2D flux map. The residual from the quasar signature is visible in the residual map in the top-right panel of the figure. Because of the nature of the object, the kinematics are not well constrained in this case. However, the PA is well constrained to be equal to 90 degrees. We derive 2 $V_{\rm max}$ sin i $\sim$27km/s, corresponding to an inclination-corrected maximum velocity of $V_{\rm max}$$\sim$20km/s, a dispersion $\sigma \sim$60 km/s and the \ha\  half-light radius r$_{1/2}=$0.2" ($\sim$2kpc). This latter value is used with the virial theorem to calculate the total mass of the system since the rotation is not established yet (Wright et al. 2009):

\begin{equation}
M_{\rm dyn} = C \sigma^2~r_{1/2}~/~G
\end{equation}

where C is a constant (C=5 for a spherically uniform density profile). We find M$_{\rm dyn}$=10$^{9.8}$ $M_{\odot}$ which is consistent with the estimate of Krogager et al. (2013) of M$_{\rm dyn}$=10$^{9.4}$ $M_{\odot}$ based on 1D emission lines measurements. Following the procedure described for Q0452$-$1640, we compute the mass of gas in this system to be M$_{\rm gas}$=10$^{9.7}$ $M_{\odot}$, i.e. significantly larger than the gas mass of M$_{\rm gas}$=10$^{9.0}$ $M_{\odot}$ derived by Krogager et al. (2013). The difference arises because our estimate of the SFR is larger than Krogager et al. Finally, Krogager et al. (2013) have used HST images to constrain the stellar mass of the system using spectral energy distribution (SED) fitting. They derive M$_{*}$=10$^{9.3}$ $M_{\odot}$. The dynamical mass of this system is lower than in other objects of this study and comparable to the gas and stellar masses. All these values are listed in Table~\ref{t:kine}.

\subsection{Q2352$-$0028, \zabs=1.0318}

The sub-DLA towards this quasar has a \hi\ column density \lognhi=19.81$^{+0.14}_{-0.11}$. The absorbing-galaxy was first reported by 
P\'eroux et al. (2012) using low-resolution SINFONI data. Unfortunately, the new high-resolution SINFONI data are not centered on the targeted galaxy as requested so that the field-of-view position varies from one Observing Block to another. As a result, only a small fraction of the total exposure time is actually spent on the target. Another consequence is that the quasar is not always included in the field, thus complicating the co-addition of the individual cubes. We thus find more suitable to use the 0.25-arcsec pixel-scale resolution data to perform the dynamical analysis described below. 

\begin{figure*}
\begin{center}
\includegraphics[height=4.5cm, width=5.5cm, angle=0]{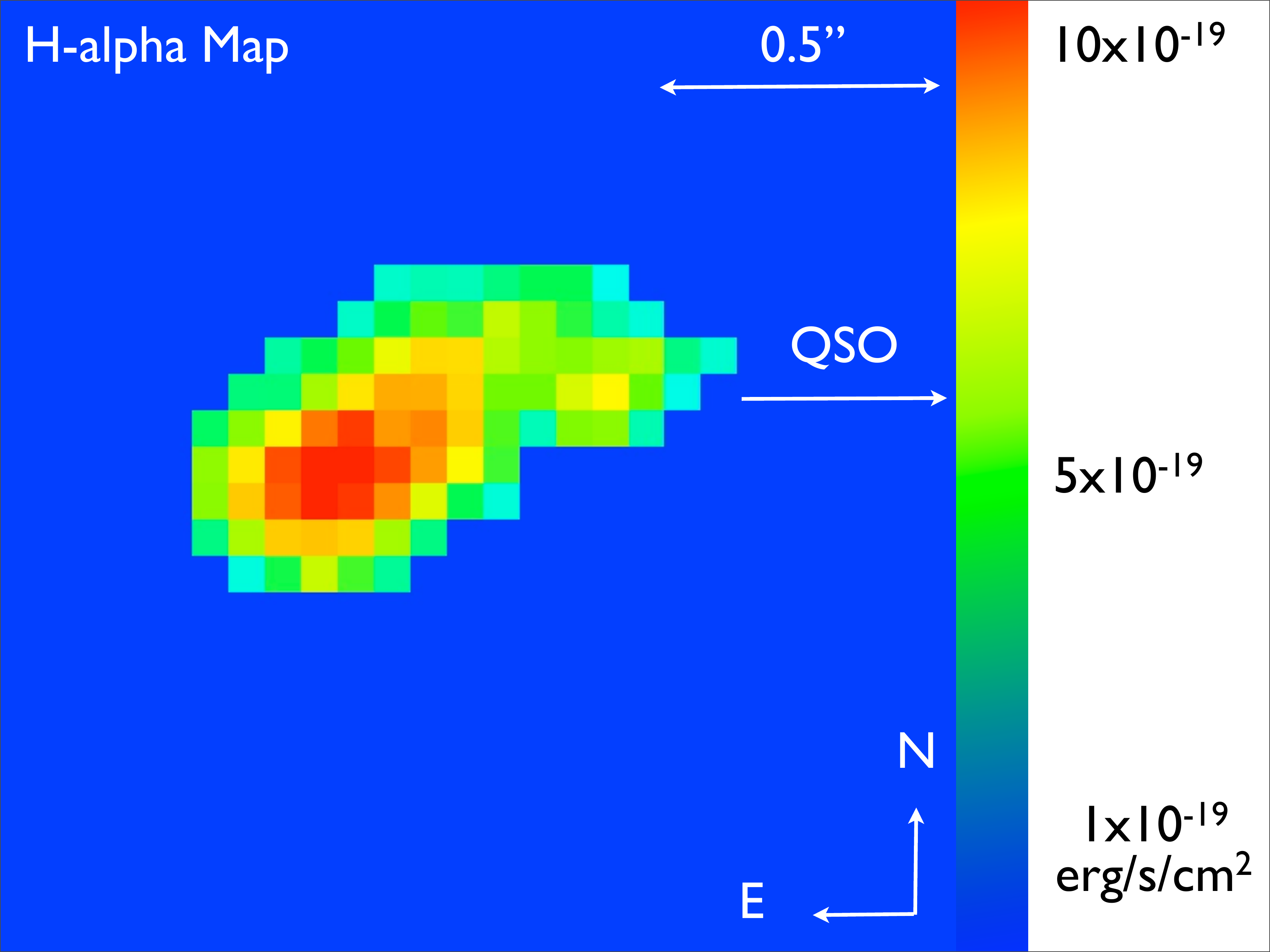}
\includegraphics[height=4.5cm, width=5.5cm, angle=0]{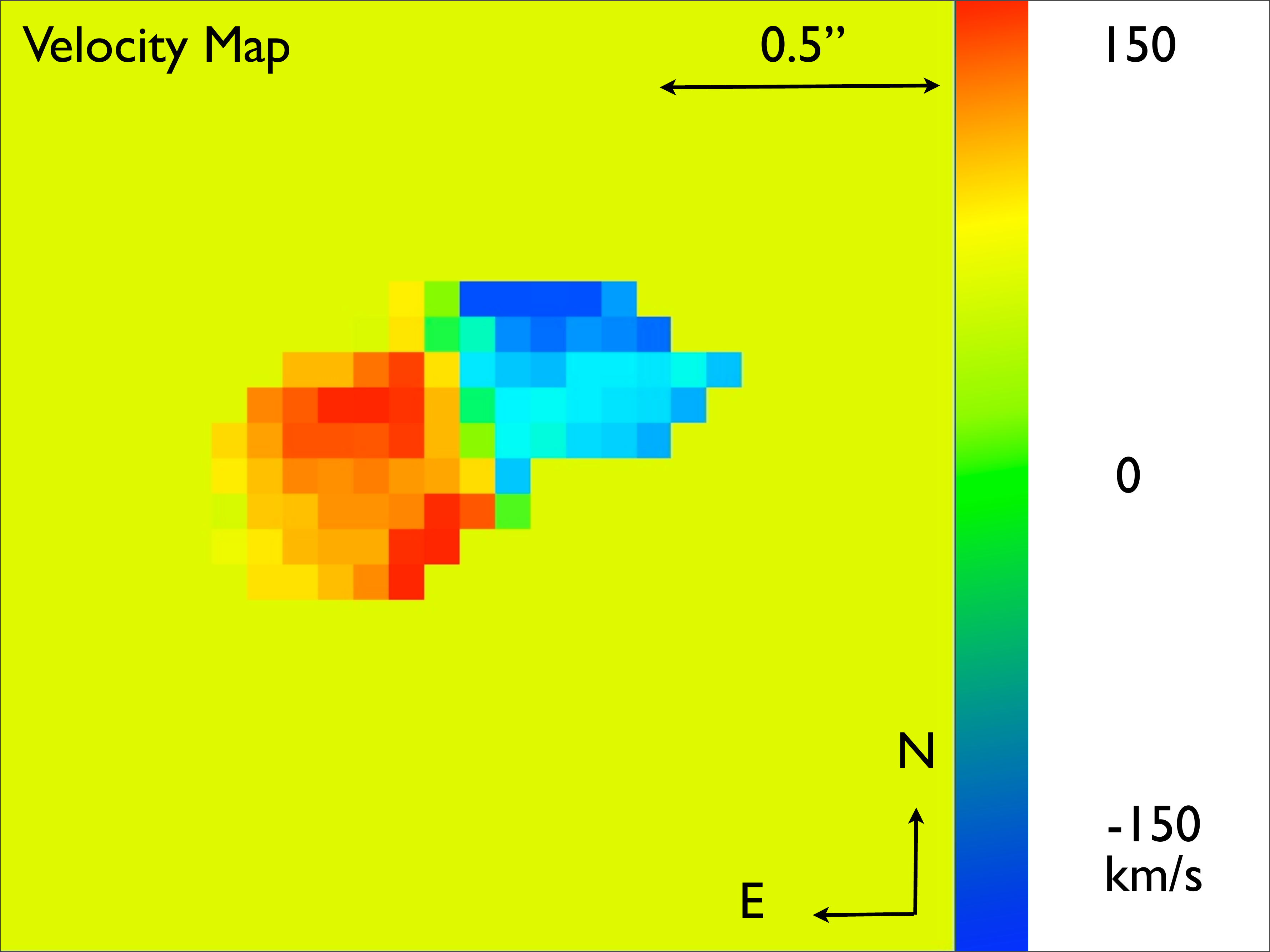}
\includegraphics[height=4.5cm, width=5.5cm, angle=0]{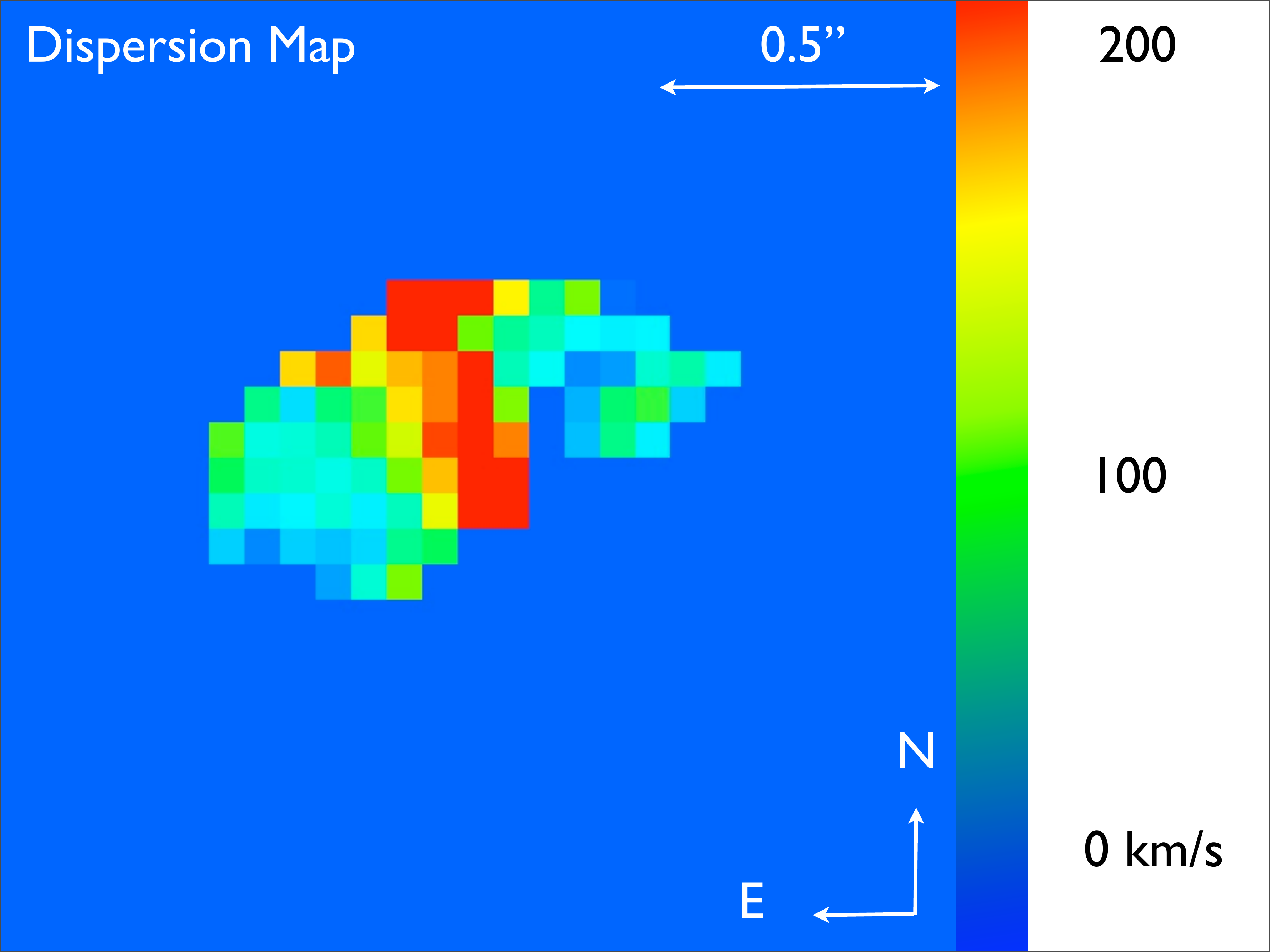}
\caption{{\bf \ha\ flux map, \ha\ velocity field and \ha\ velocity dispersion maps of the sub-DLA-galaxy towards Q2352$-$0028, \zabs=1.0318.} The velocity map indicates a strong gradient as expected from a rotating disc.}
\label{f:Q2352_kine}
\end{center}
\end{figure*}

\begin{figure*}
\begin{center}
\includegraphics[height=13cm, width=15cm, angle=0]{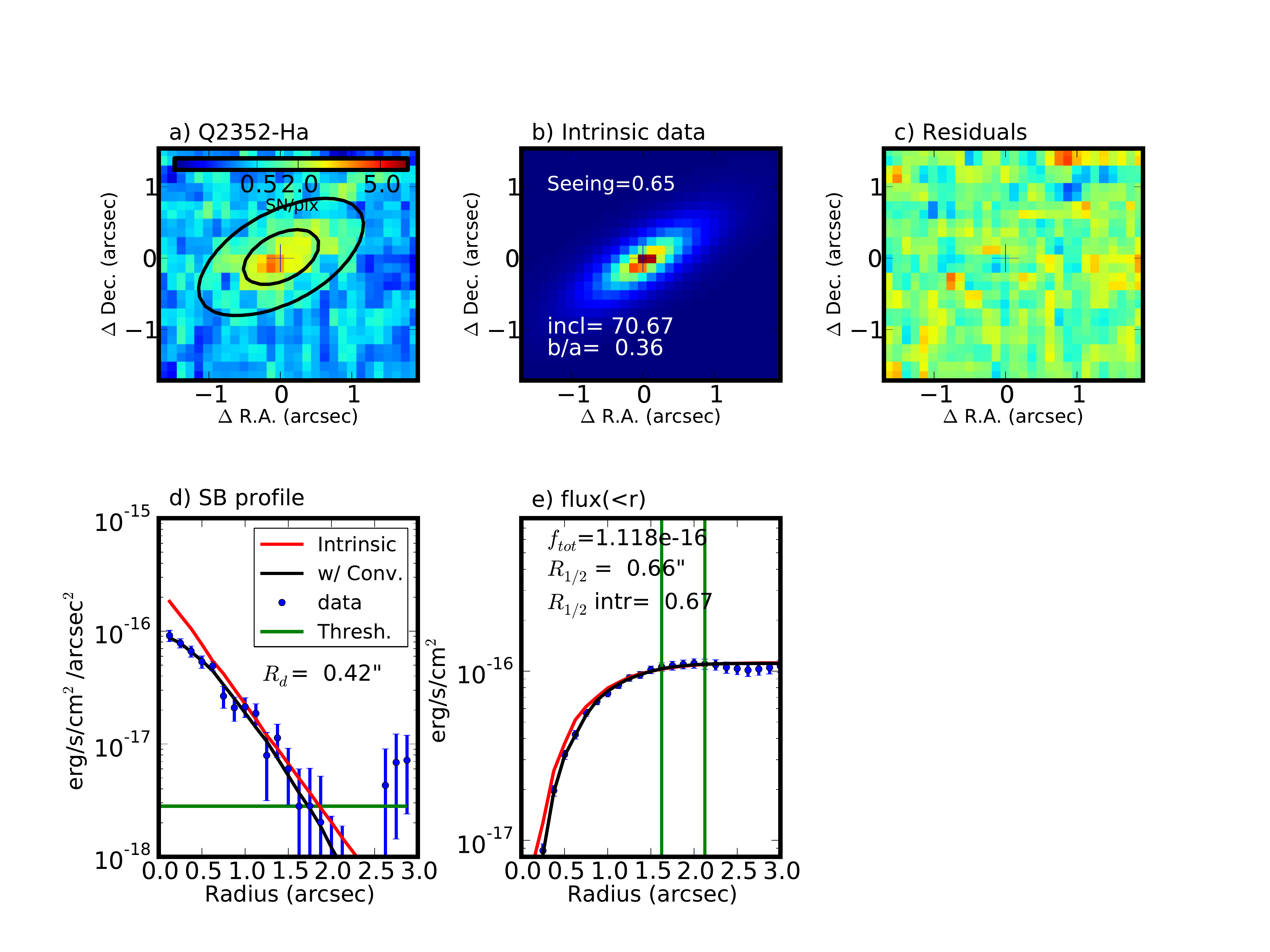}
\caption{{\bf Results from the fit to the 2D flux map of the absorber in the field of Q2352$-$0028 at \zabs=1.0318.}  In that case, the fit converges and the findings are particularly robust. Indeed, the residual map does not show any particular feature. The galaxy is elongated with an inclination sin i=0.93. We derive a high value for the maximal velocity $V_{\rm max}\sim$140 km/s and a dispersion of $\sigma \sim$120 km/s peaking at the dynamical centre. Similarly, the \ha\  half-light radius is well constrained to be r$_{1/2}=$0.7".   }
\label{f:Q2352_3Dfit}
\end{center}
\end{figure*}

Figure~\ref{f:Q2352_kine} presents the \ha\ flux map, \ha\ velocity field and \ha\ velocity dispersion maps of the absorbing-galaxy detected towards Q2352$-$0028. The velocity map indicates a strong gradient as expected from a rotating disc. Similarly, the dispersion is found to peak at the centre of the object. In this case, the 3D kinematical fit converges and the findings are particularly robust. Figure~\ref{f:Q2352_3Dfit} illustrates the results of the fit to the 2D flux map. The residual map does not show any particular feature, indicating a good fit. The galaxy is elongated with an inclination sin i=0.93. The global velocity dispersion for this galaxy is larger than the one from rotationally supported galaxies at these redshifts which ranges from $\sigma$$\sim$50--100km/s.
We derive a high value for the maximal velocity 2 $V_{\rm max}$ sin i $\sim$260km/s, corresponding to an inclination-corrected value of $V_{\rm max}$$\sim$140km/s and a dispersion of $\sigma$$\sim$120km/s peaking at the dynamical centre. Similarly, the \ha\ half-light radius is well constrained to be r$_{1/2}=$0.7" ($\sim$5kpc). We use the same procedure as outlined above to estimate the various values of mass and find M$_{\rm dyn}$=10$^{10.4}$ $M_{\odot}$. The mass of gas in this system is small, M$_{\rm gas}$=10$^{8.8}$ $M_{\odot}$. In this case, the gas mass is significantly lower than the dynamical mass, indicating a very low gas fraction in the object. Finally, we derive the mass of the halo in which this system resides to be M$_{\rm halo}$=10$^{11.8}$ $M_{\odot}$. All these values are listed in Table~\ref{t:kine}.

\section{Gas Flows}

The five quasar lines-of-sight presented here and in previous publications (see Table~\ref{t:kine}) offer probes of the gas surrounding the galaxies detected in emission. These extended gaseous haloes could be to either due to i) galaxy group environments producing tidal streams and stripped gas, ii) outflowing gas, iii) infalling gas, or iv) gas co-rotating with the disk of the galaxy. Out of the five cases presented in this work, two of the objects (Q0302$-$223 and Q0452$-$1640) are known to be interacting systems and the gas seen in absorption might well be due to an environment producing tidal streams and stripped gas from these galaxies. The environment of the remaining systems is not well constrained given the limited field-of-view of our SINFONI observations. Other lines of evidence, detailed below, can help constraining the nature of the gas probed in absorption.

\begin{figure*}
\begin{center}
\includegraphics[height=16.cm, width=7.cm, angle=-90]{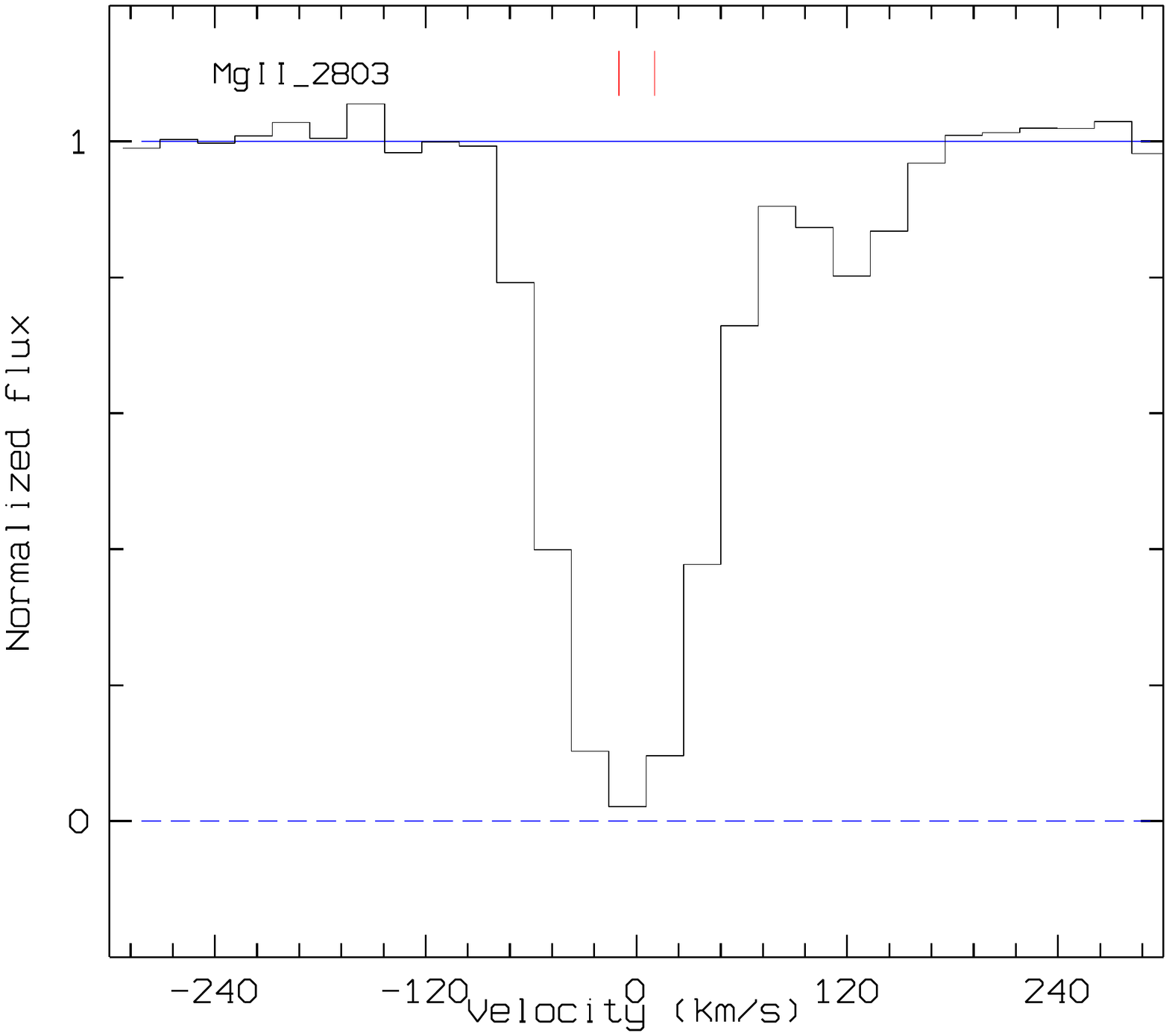}
\includegraphics[height=16.cm, width=7.cm, angle=-90]{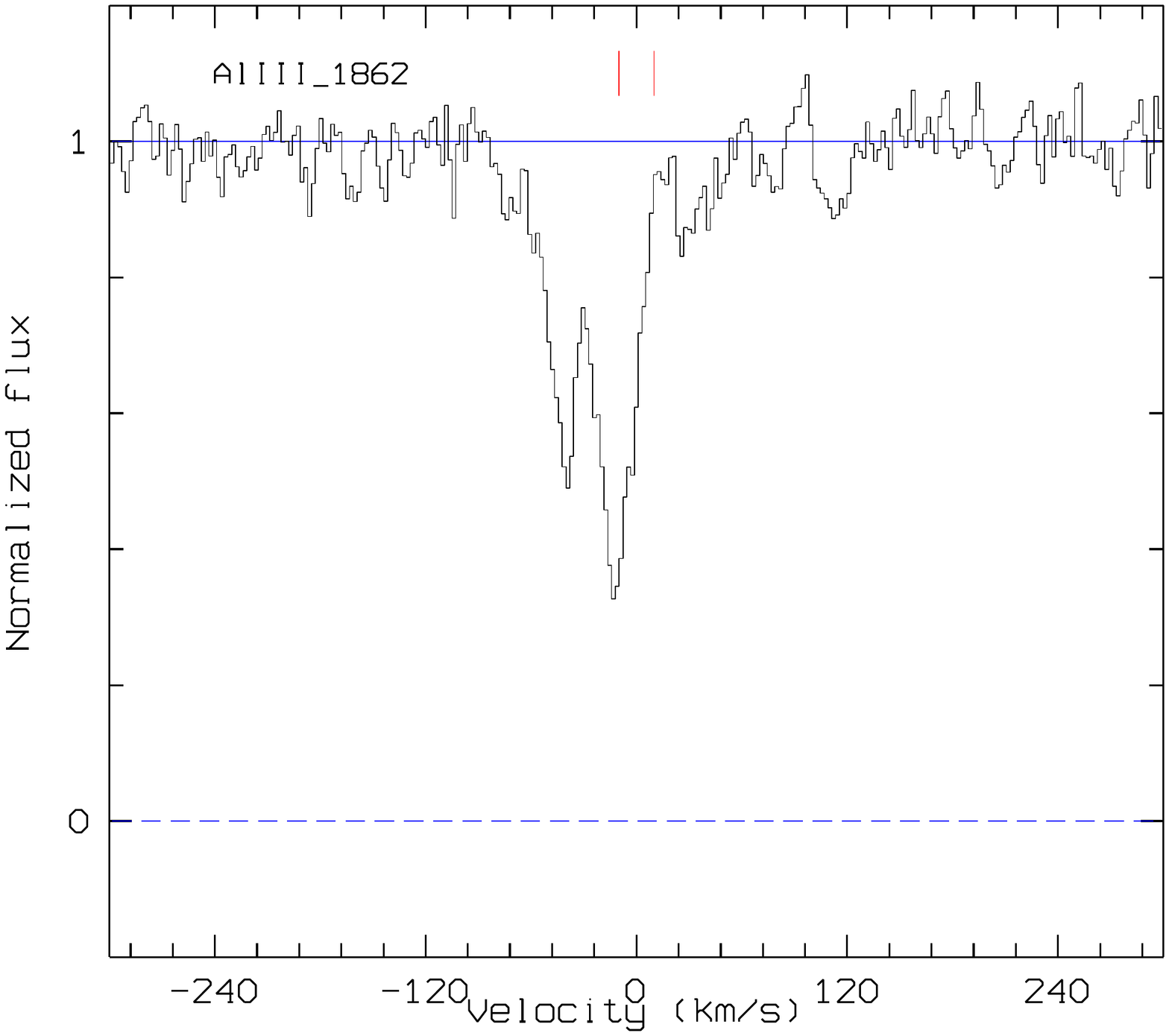}
\includegraphics[height=16.cm, width=7.cm, angle=-90]{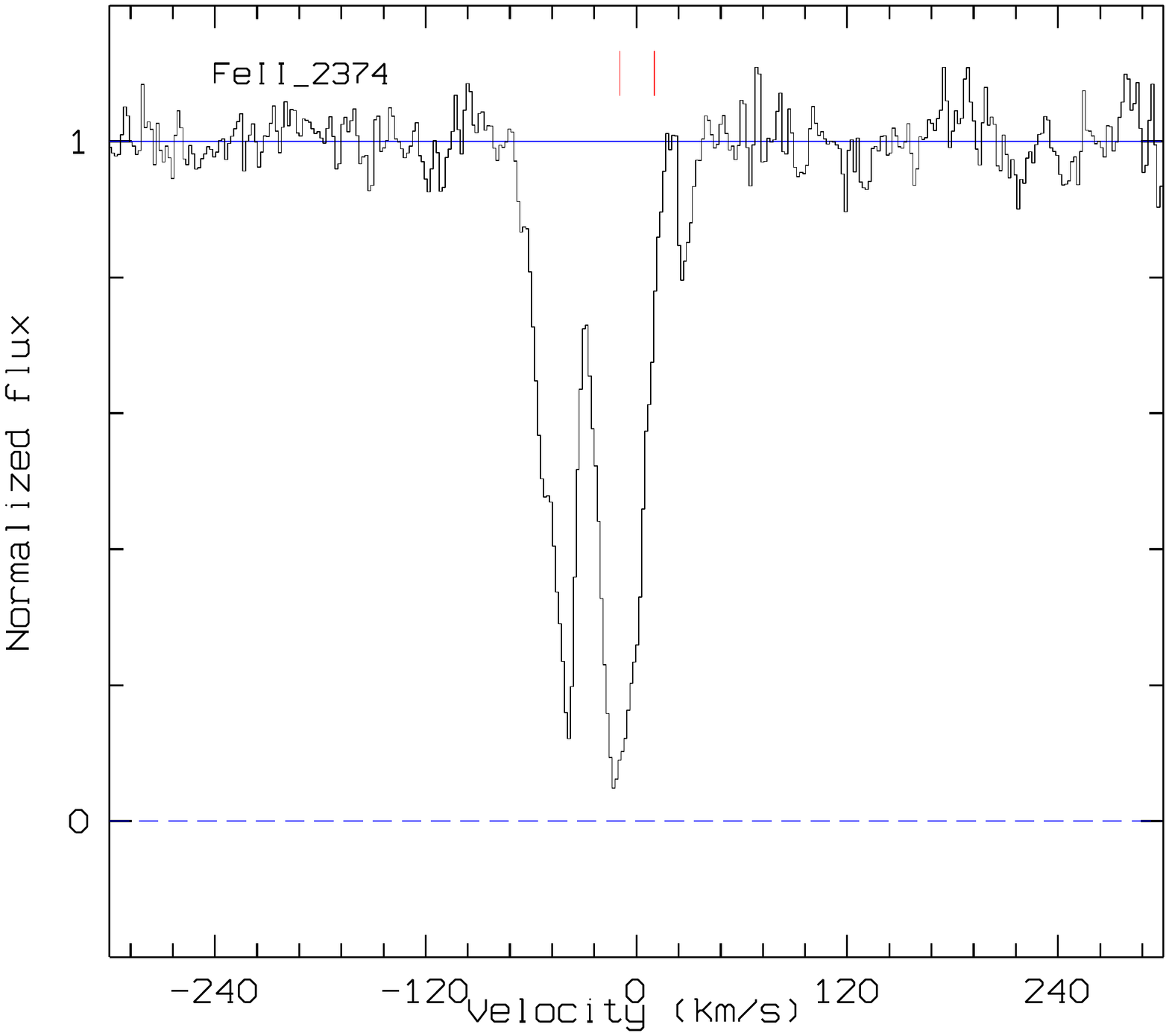}
\caption{{\bf Absorption profiles towards Q0302$-$223 seen along the line-of-sight to the background quasar.} In this and the following, we plot the low-ionisation absorption profiles of \feii\ $\lambda$ 2374 and  \mgii\ $\lambda$ 2803 as well as the intermediate-ionisation profiles of \aliii\ $ \lambda$ 1862. The red tick marks indicate the maximal circular velocity either side of the galaxy's systemic redshift. In this case, the \feii\ and \aliii\ data are from Keck/HIRES high-resolution spectroscopy while the \mgii\ profile is from the medium-resolution X-Shooter spectrum of P\'eroux et al. (2013b). }
\label{f:Q0302_XSH}
\end{center}
\end{figure*}

\begin{figure*}
\begin{center}
\includegraphics[height=16.cm, width=7.cm, angle=-90]{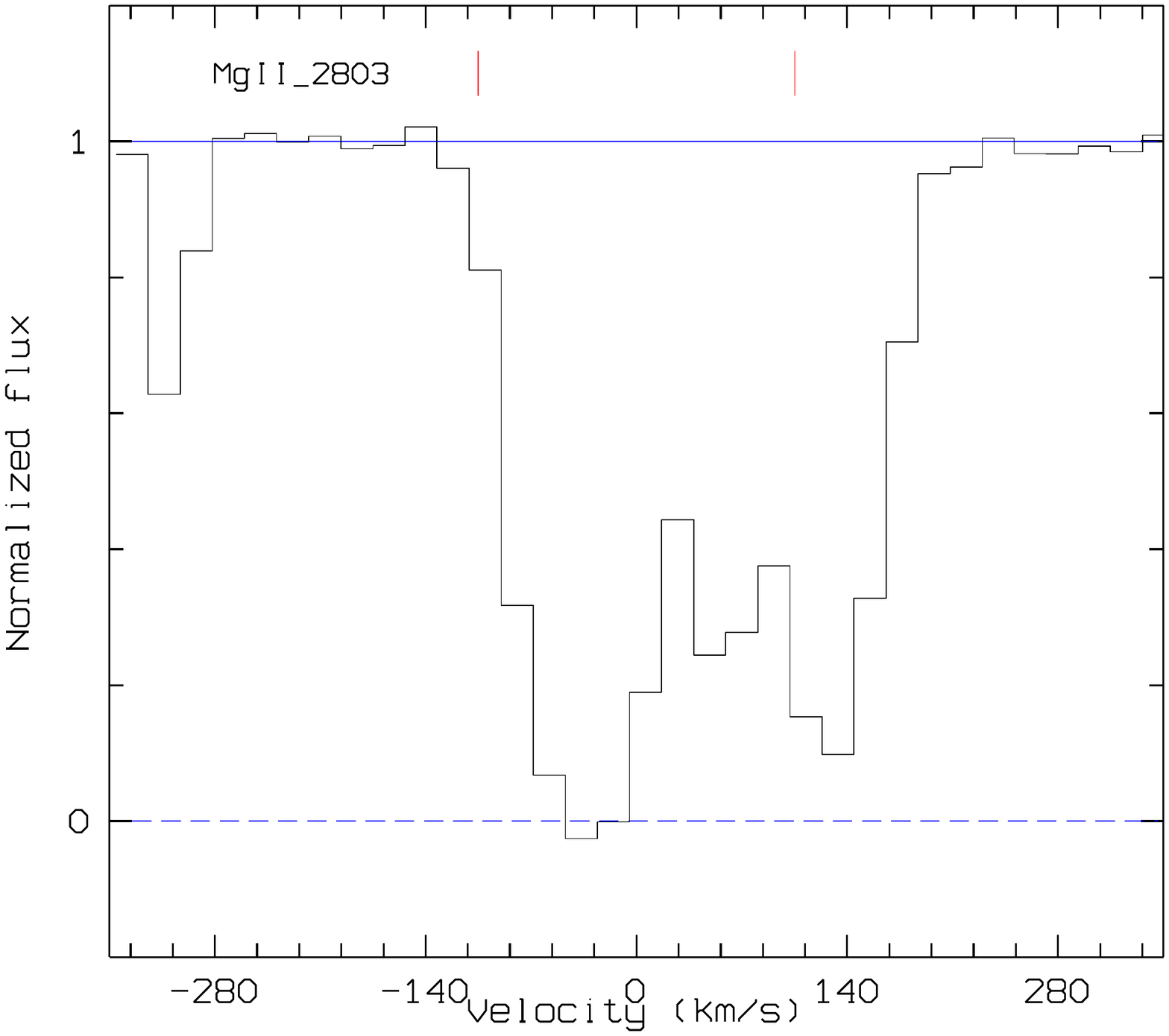}
\includegraphics[height=16.cm, width=7.cm, angle=-90]{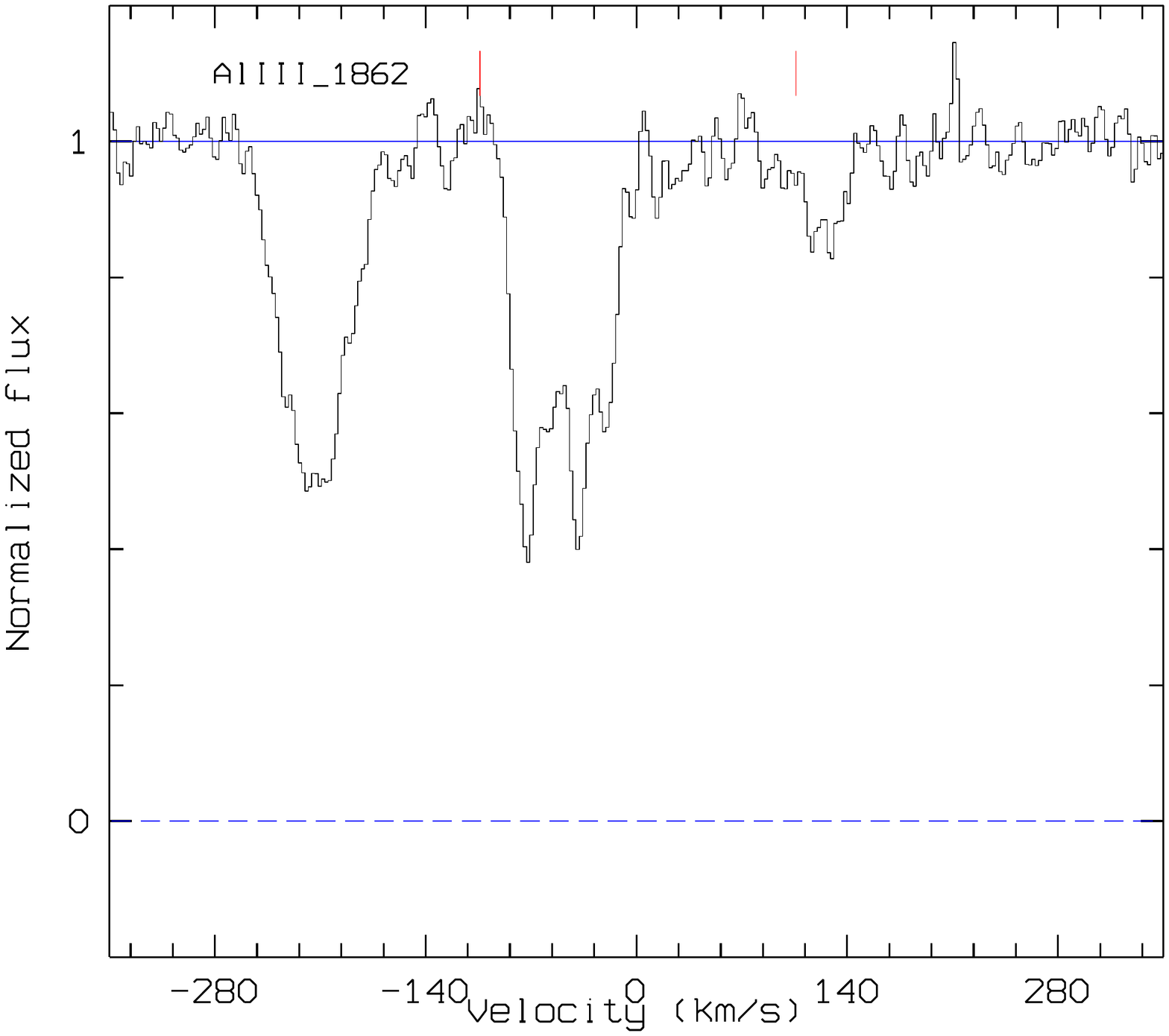}
\includegraphics[height=16.cm, width=7.cm, angle=-90]{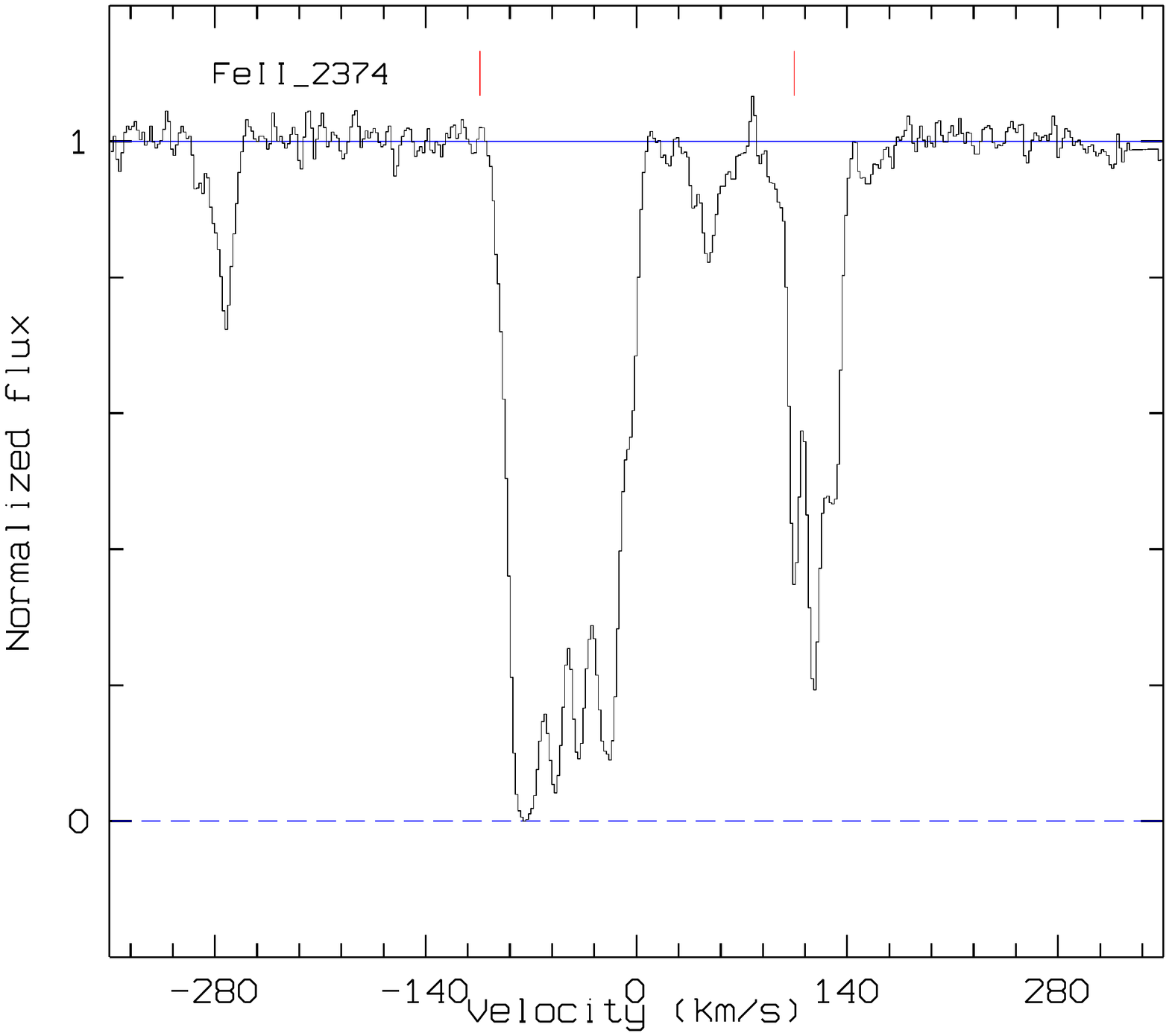}
\caption{{\bf Absorption profiles towards Q0452$-$1640 seen along the line-of-sight to the background quasar.} In this case, the \feii\ and \aliii\ data are from VLT/UVES high-resolution spectroscopy while the \mgii\ profile is from the medium-resolution X-Shooter spectrum of P\'eroux et al. (2013b). The feature at $-$250 km/s seen in the \aliii\ $\lambda$ 1862 panel is a blend from the \lya\ forest.}
\label{f:Q0452_XSH}
\end{center}
\end{figure*}

\begin{figure*}
\begin{center}
\includegraphics[height=16.cm, width=7.cm, angle=-90]{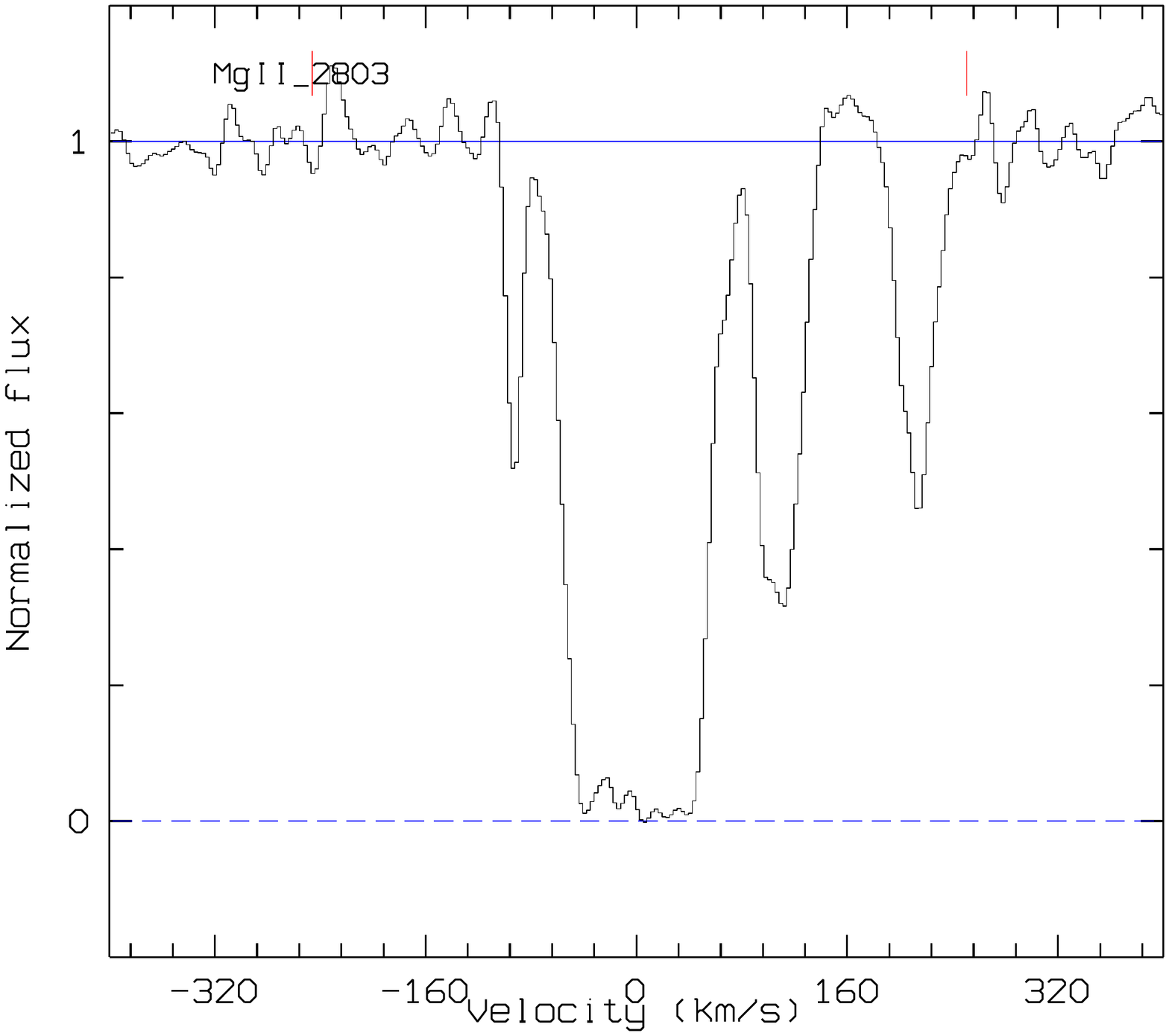}
\includegraphics[height=16.cm, width=7.cm, angle=-90]{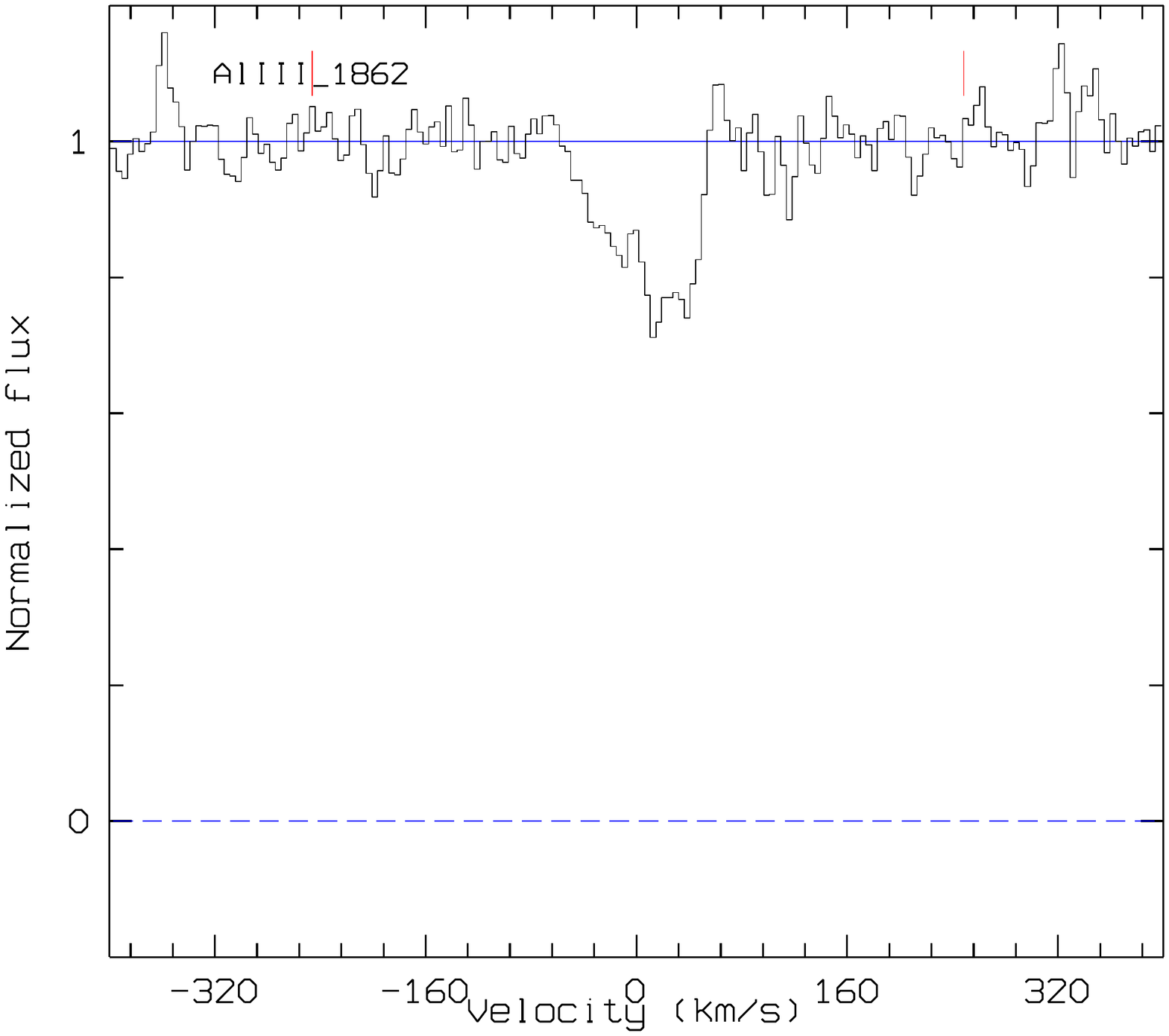}
\includegraphics[height=16.cm, width=7.cm, angle=-90]{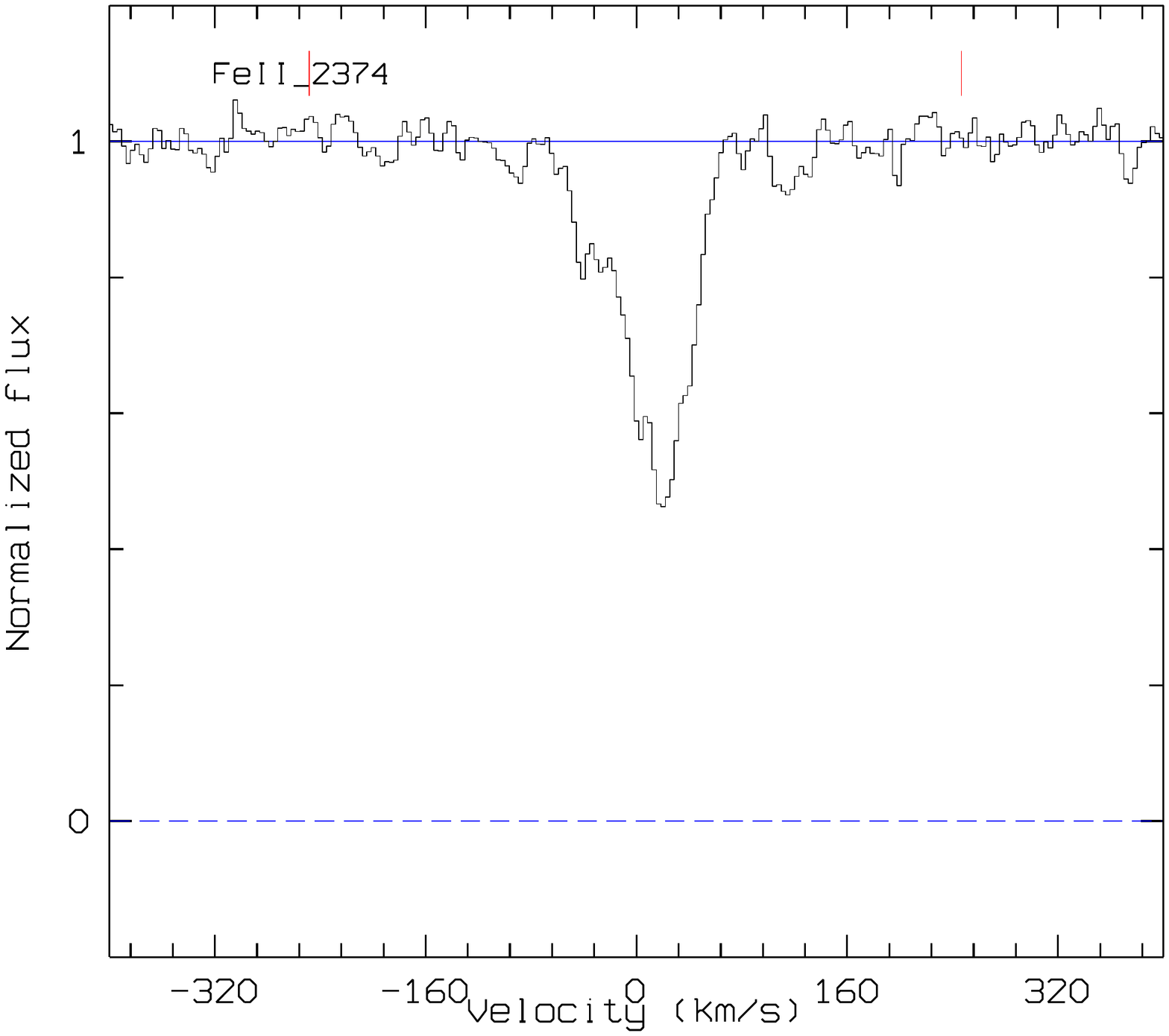}
\caption{{\bf Absorption profiles towards Q1009$-$0026 seen along the line-of-sight to the background quasar.} In this case, all the lines shown are from Magellan/MIKE high-resolution spectroscopy.}
\label{f:Q1009_XSH}
\end{center}
\end{figure*}

\begin{figure*}
\begin{center}
\includegraphics[height=16.cm, width=7.cm, angle=-90]{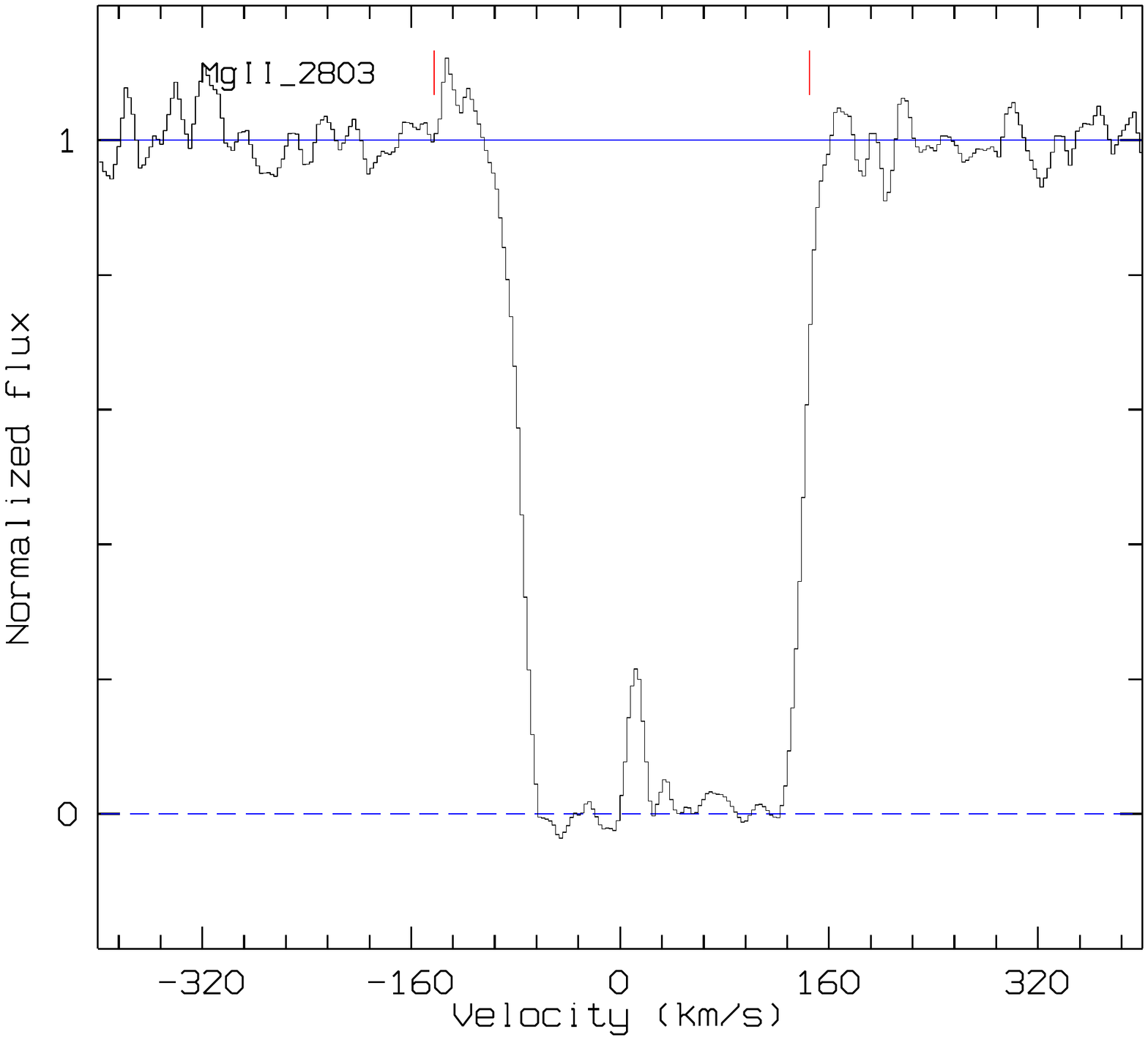}
\includegraphics[height=16.cm, width=7.cm, angle=-90]{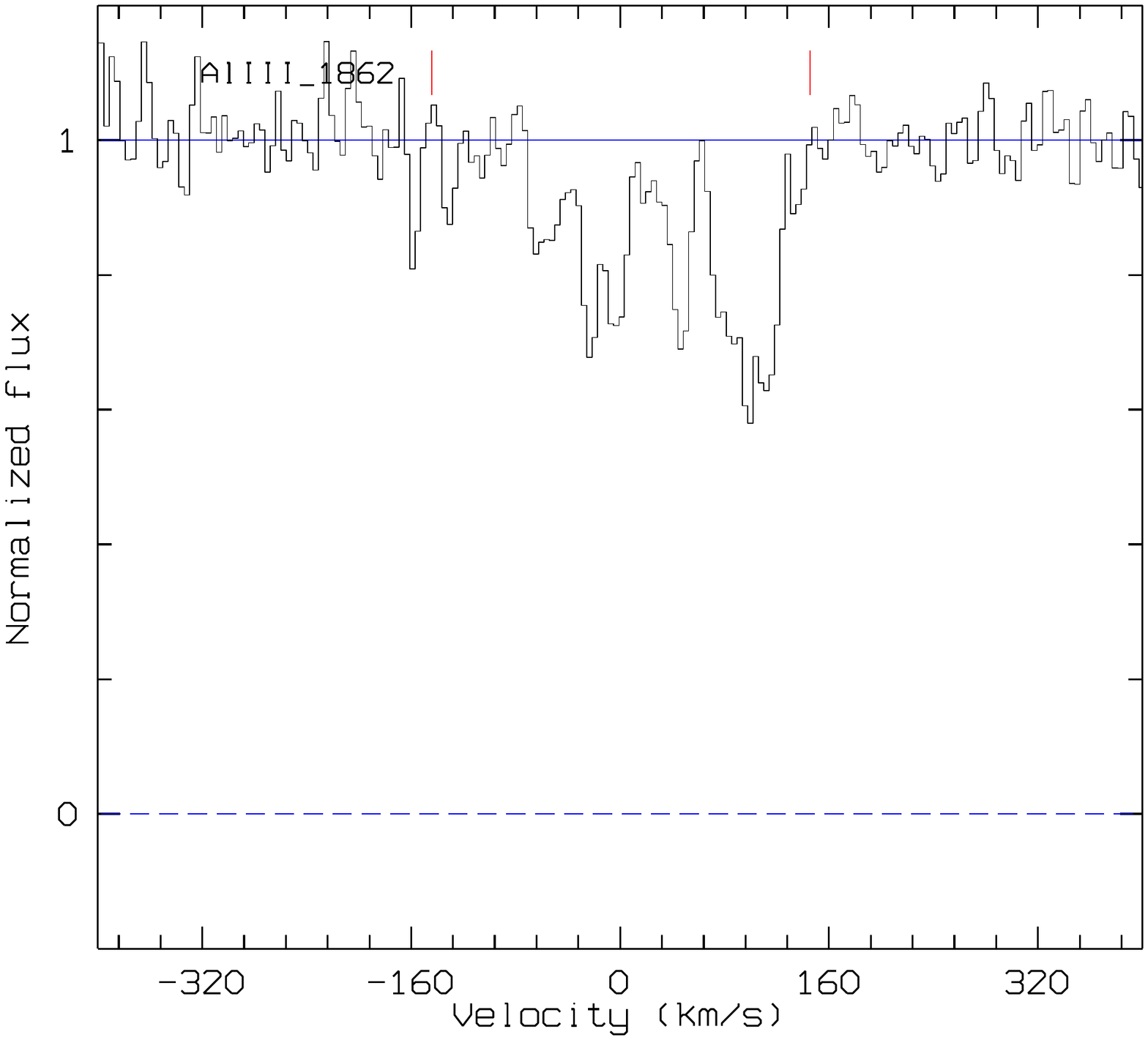}
\includegraphics[height=16.cm, width=7.cm, angle=-90]{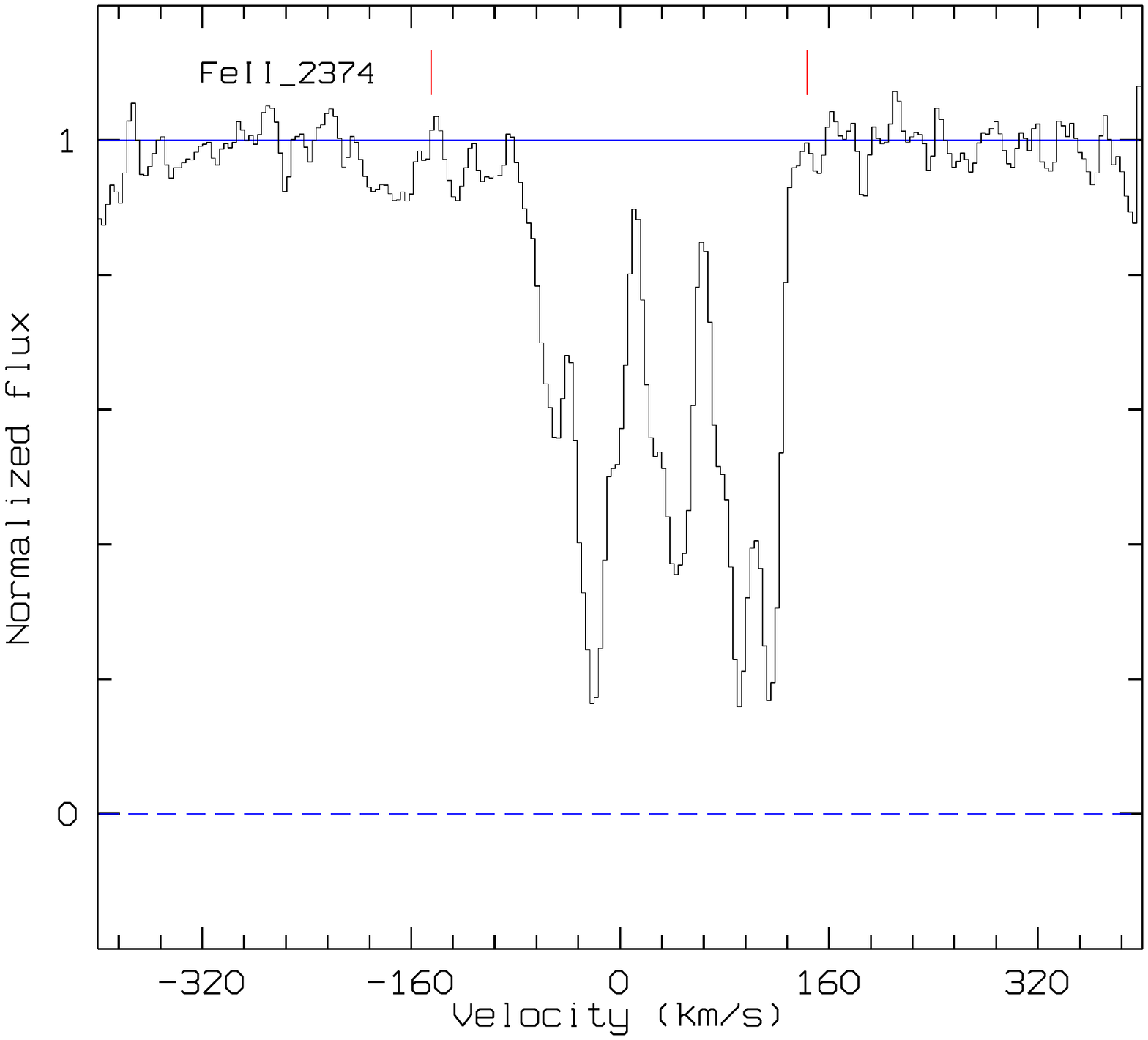}
\caption{ {\bf Absorption profiles towards Q2352$-$00286 seen along the line-of-sight to the background quasar.} In this case, all the lines shown are from Magellan/MIKE high-resolution spectroscopy.}
\label{f:Q2352_XSH}
\end{center}
\end{figure*}

\subsection{Star Formation per Unit Area}

Both observations  from Heckman (2002, 2003) and models (Murray, M\'enard \& Thompson 2011; Kornei et al. 2012) argue that the star formation per unit area, $\Sigma_{\rm SFR}$, is one of the most relevant indicators of galaxy outflows. These outflows are ubiquitous in galaxies where the global star-formation rate per unit area exceeds $\Sigma_{\rm SFR}$=0.1 M$_{\odot}$/yr/kpc$^2$ (Heckman 2002). This criterion applies to local starbursts and even high-redshift Lyman Break galaxies. In our small sample of five objects, all the systems have  $\Sigma_{\rm SFR}>$0.1 M$_{\odot}$/yr/kpc$^2$ (see Table~\ref{t:kine}). In other words, all these objects might produce winds, especially the compact galaxy towards Q2222$-$0946.

\subsection{Absorption Profiles}

\noindent

\underline{Emission/Absorption Kinematics:} the dynamical information described above can be further compared with the kinematical properties of the neutral gas associated with these systems and seen in absorption along the line-of-sight to the background quasars. Such information is rarely available for a given system, but for our small sample of five high-\nhi-absorbers presented here and in P\'eroux et al. (2011b), high-resolution spectroscopy of the background quasar is available to us. In Figures~\ref{f:Q0302_XSH} to \ref{f:Q2352_XSH}, we show the low-ionisation absorption profiles of \feii\ $\lambda$ 2374 and intermediate-ionisation profiles of \aliii\ $ \lambda$ 1862 from Keck/HIRES, VLT/UVES or Magellan/MIKE observations. The low-ionisation states, such as \mgi, \mgii, \sii, \cii\ and \ciii\ are indeed ideal for tracing cold mode accretion ($<$100,000K; Stewart et al. 2011) given the metallicity, temperatures and densities expected. In addition, the \mgii\ $\lambda$ 2803 is plotted from either these data when it is covered or from our recent X-Shooter data (P\'eroux et al. 2013b) when available. 

In the case of Q0302$-$223, shown in Figure~\ref{f:Q0302_XSH}, the \feii\ line of $\lambda$ 2374, typical of the unsaturated profile in this system, is composed of two main components separated by 36 km/s (Pettini et al. 2000) precisely centered on the galaxy systemic redshift. The \mgii\ doublet falls in an UVES spectral gap so we show the \mgii\ $\lambda$ 2803 profile from the lower-resolution X-Shooter data (P\'eroux et al. 2013b). Additional weaker components are observed at v=35 and 121 km/s and are more obvious in stronger \feii\ lines. These values are larger than the maximal velocity $V_{\rm max}\sim$11km/s derived from the stellar emission in the SINFONI data reported in P\'eroux et al. (2011b) and indicate that additional processes have to be invoked to explain the gas seen in absorption along the line-of-sight to the quasar at impact parameter of b=25 kpc. 
 
In the case of Q0452$-$1640, shown in Figure~\ref{f:Q0452_XSH}, the absorption profile shows strong components either side of the galaxy systemic redshift (P\'eroux et al. 2008). The high-quality of the UVES spectrum means that 18 components are required to fit the profile. The full structure spans $\Delta$v=230 km/s. However, the \mgii\ doublet falls in an UVES spectral gap so we show the MgII $\lambda$ 2803 profile from the lower-resolution X-Shooter data (P\'eroux et al. 2013b). The extent of the gas probed in absorption is therefore comparable with the maximal velocity seen in the stellar component of the galaxy at the top of the complex reported in this paper, which is derived to be $V_{\rm max}\sim$100km/s. This seems to suggest that the gas probed in absorption is due to either a bipolar outflow probed at an impact parameter of b=16 kpc or the merging process of the two structures detected with SINFONI in emission.

In the case of Q1009$-$0026, shown in Figure~\ref{f:Q1009_XSH}, the absorption profile is highly asymmetrical around the galaxy systemic redshift. For this absorber, Meiring et al. (2007) have used seven components to fit the low-ionisation line profiles over more than $\Delta$v=334 km/s. This profile includes a strong component at negative velocities and several smaller components at positive velocities. The gas probed in absorption is therefore comparable to or smaller than the maximal velocity seen in the stellar component of the galaxy reported in P\'eroux et al. (2011b), which is derived to be $V_{\rm max}\sim$250km/s at an impact parameter of b=39 kpc. The totality of the gas seen in absorption might therefore be explained by a combination of rotation and local dispersion (Cohn \& York 1982; Cowie \& York 1978) but the asymmetrical absorption profile is suggestive of wind (Bouch\'e et al. 2012).

For the absorbing galaxy towards Q2222$-$0946, Krogager et al. (2013) have used a lower-resolution X-Shooter spectrum to fit five components to the low-ionisation line profiles over more than $\Delta$v$\sim$200 km/s. This velocity is considerably larger than the maximal velocity $V_{\rm max}\sim$20km/s derived from the stellar emission in the SINFONI data reported in this paper. This large difference strongly suggests that the gas seen in absorption is not due to the co-rotating material around the absorbing-galaxy but that additional physical processes such as an outflow are required even at a small impact parameter of b=6 kpc. Note that this object is the only z$\sim$2 system in the sample and therefore it is possible that the SINFONI observations are limited by the low surface brightness of the outskirt of the system.

In the case of Q2352$-$0028, shown in Figure~\ref{f:Q2352_XSH}, the absorption profile of the unsaturated ions such as \feii\ $\lambda$ 2374 are fitted with 11 components over 
$\Delta$v$\sim$220 km/s either side of the systemic redshift of the absorbing galaxy (i.e. $\Delta$v$\sim$110 km/s with respect to v=0). This value is slightly less than the $V_{\rm max}\sim$140km/s derived from the stellar emission at an impact parameter of b=12 kpc reported in this paper, which is difficult to reconcile with the hypothesis of wind or infall. 

\vspace{0.5cm}

\underline{\mgii\ Kinematics:} Various authors (Bouch\'e et al. 2007a; Kacprzak et al. 2011b; Gauthier \& Chen 2012) have argued that \mgii\ with large equivalent widths (EW) are a signature of winds. This is consistent with Kacprzak et al. (2011b) who showed that absorption strength is
correlated with the orientation of the galaxy major axis, implying that a significant fraction of weaker \mgii\ absorption systems are
more likely accreting towards the galaxy via cold flows. Indeed, the authors argue that \mgii\ systems at intermediate-redshift with EW(\mgii\ $\lambda$ 2796)$<$1\AA\ are not expected to be produced by star formation driven winds. 

Here, we have access to the \mgii\ profile of each of the five systems under study, allowing us to further probe the gas flows around these galaxies. For Q0302$-$223 and Q0452$-$1640, we use our X-Shooter data (P\'eroux et al. 2013b). For the first absorber, we measure EW(\mgii\ $\lambda$ 2796)=2.118\AA\ and EW(\mgii\ $\lambda$ 2803)=1.875\AA\ in line with EW(\mgii\ $\lambda$ 2796)=2.34\AA\ and EW(\mgii\ $\lambda$ 2803)=1.92\AA\ reported by Petitjean \& Bergeron (1990) from lower resolution data. For the second absorber, we measure an EW(\mgii\ $\lambda$ 2796)=4.869\AA\ and EW(\mgii\ $\lambda$ 2803)=4.373\AA. Clearly, these lines are heavily saturated at the X-Shooter resolution (see Figure~\ref{f:Q0302_XSH} and \ref{f:Q0452_XSH}). For Q1009$-$0026, Meiring et al. (2007) report EW(\mgii\ $\lambda$ 2796)=1.792\AA\ and EW(\mgii\ $\lambda$ 2803)=1.508\AA. For Q2222$-$0946, the \mgii\ doublet is covered by the X-Shooter data of Krogager et al. (2013) but these authors do not report any EW. We therefore use the low-resolution Sloan spectrum to derive: EW(\mgii\ $\lambda$ 2796)=4.596\AA\ and EW(\mgii\ $\lambda$ 2803)=3.602\AA. Finally, Meiring et al. (2009) report EW(\mgii\ $\lambda$ 2796)=2.069\AA\ and EW(\mgii\ $\lambda$ 2803)=1.984\AA\ in the spectrum of Q2352$-$0028. Therefore in all the five cases studied here, we measure EW(\mgii\ $\lambda$ 2796) $>$ 1\AA\ which is believed to be a signature of winds.

Previous observations of Steidel et al. (2002) and Kacprzak et al. (2010) show that \mgii\ absorption
residing fully to one side of the galaxy systemic velocity is usually aligned with expected galaxy rotation direction, with the
absorption essentially mimicking the extension of the galaxy rotation curve out into the halo. We do not find evidence of this in any of the five systems. Findings based on $\Lambda$CDM simulations (Kacprzak et al. 2010) also claim that a saturated \mgii\ absorption profile spanning both sides of the
galaxy systemic velocity may be a direct signature of outflows. In all five cases presented here, the \mgii\ $\lambda$ doublets are heavily saturated and span either sides of the absorbing-galaxies systemic redshifts. This suggests therefore that the gas probed in absorption towards the absorbing-galaxies can be reproduced by models of galactic winds.

\subsection{Inclination/Orientation}

Recently, Bordoloi et al. (2011) and Bouch\'e et al. (2012) argue that the \mgii\ absorption profile aligned with the minor axis of the absorbing-galaxy can be modeled by a strong bipolar wind, while those aligned with the major axis of the absorbing-galaxy are most probably due to gas associated with the disc of the galaxy. Indeed, Stewart et al. (2011) and Shen et al. (2013) argue that infall occurs along the projected galaxy major axis. This infalling gas likely produces a circumgalactic co-rotating gas component that is predominately infalling towards the galaxy and, in absorption, these structures are expected to have $\sim$100 km/s velocity offsets relative to the host galaxy and in the same direction of galaxy rotation.

We present here data of large \nhi\ absorbers which are subset of \mgii\ absorbers. Among the four cases where the system is clearly rotating (Q0302$-$223, Q0452$-$1640, Q1009$-$0026 and Q2352$-$0028), we find the quasar line-of-sight to be preferentially aligned with the major axis of the absorbing-galaxy (i.e. small angle between the galaxy major axis and the quasar line-of-sight) in two of the cases: Q0452$-$1640 and Q2352$-$0028. Therefore, based on this argument alone, the gas probed in absorption in these lines-of-sight is more likely to be either partly associated with the rotating disc of the galaxy or infall of gas. In the case of Q0302$-$223, the alignment with the minor axis is suggestive of the presence of an outflow. Finally, in the case of Q1009$-$0026, the line-of-sight to the quasar is only slightly more aligned with the minor axis of rotation of the galaxy, supportive of the presence of an outflow in this system.

\subsection{Metallicity Gradients}

\begin{figure*}
\begin{center}
\includegraphics[height=6cm, width=8cm, angle=0]{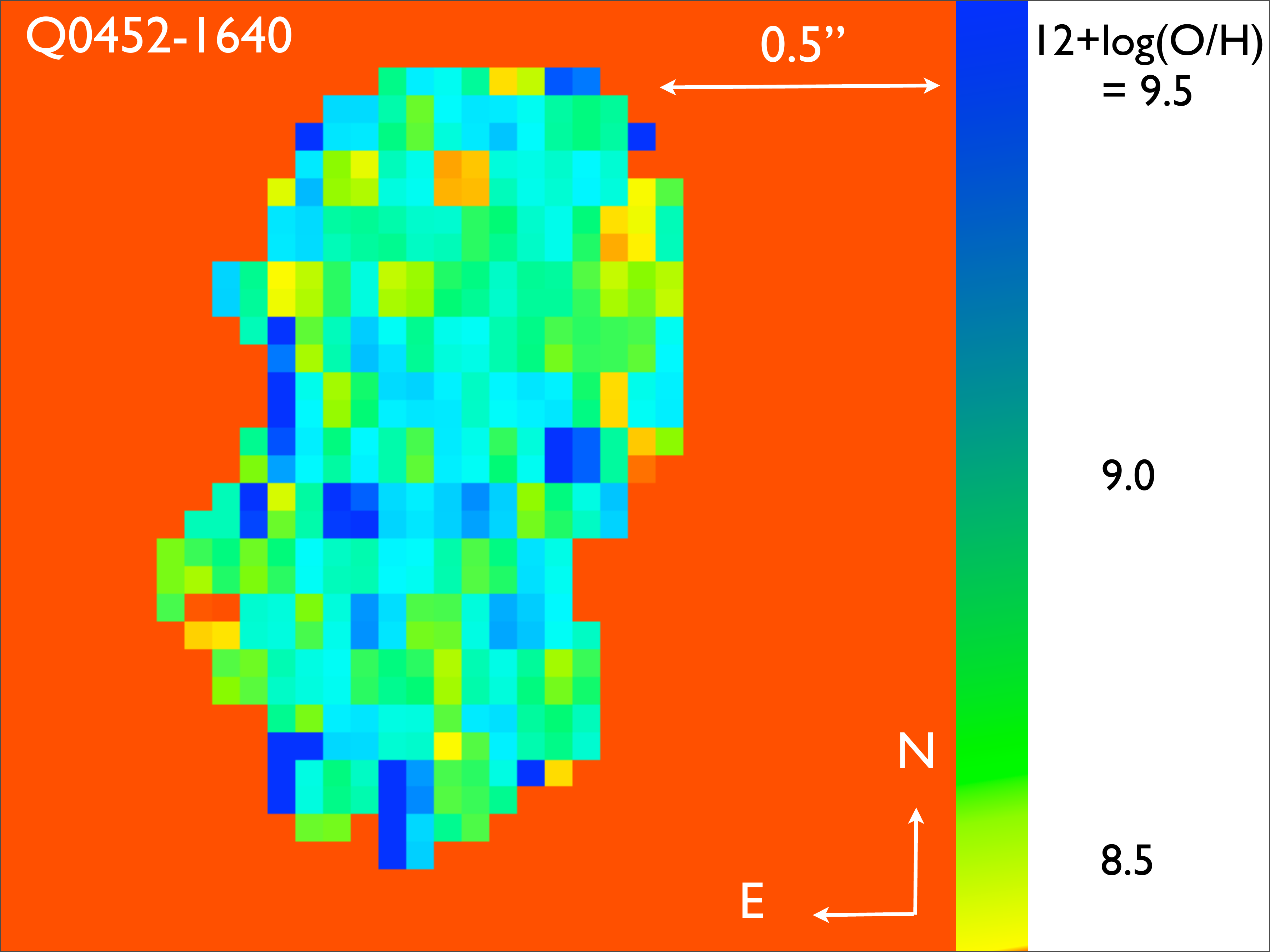}
\includegraphics[height=6cm, width=8cm, angle=0]{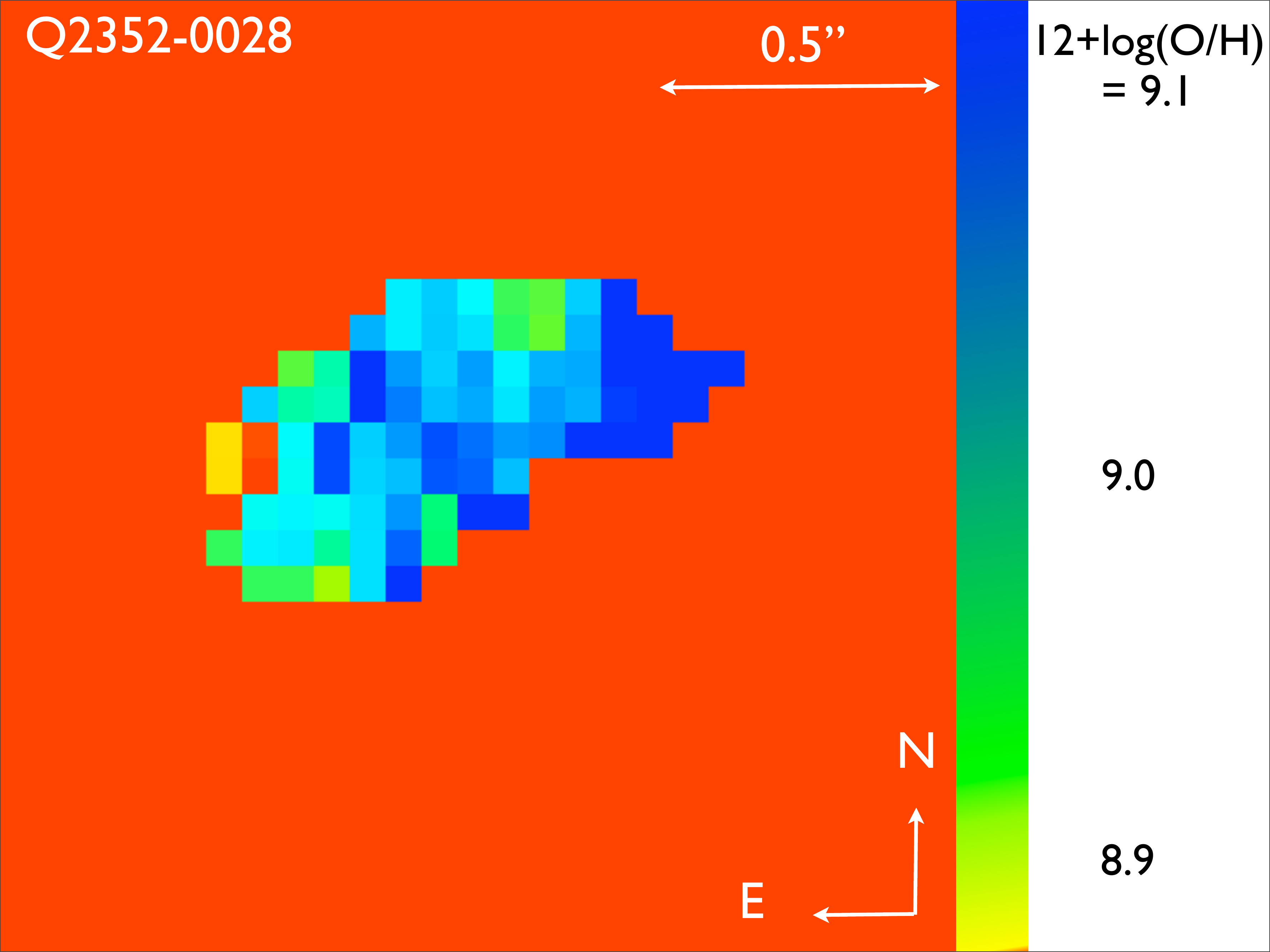}
\caption{{\bf 2D metallicity maps in units of 12+log(O/H).} The SINFONI \ha\ and \nii\ detections are used to build a 2D map of the \hii\ metallicity of the galaxies based on the N2-index. The object towards Q0452$-$1640 has a rather uniform metallicity across the extent of the galaxy. The absorbing-galaxy towards Q2352$-$0028 is showing not sign of variation over the small region covered in this object. In both cases, the gradients of metallicity with radius are only weakly negative ($-$0.11 dex/kpc and $-$0.07 dex/kpc, respectively). }
\label{f:Gradients}
\end{center}
\end{figure*}

The internal enrichment (and radial abundance gradients)
of high-redshift star-forming galaxies provides a tool for
studying the gas accretion and mass assembly process such
as gas exchange (inflows/outflows) with the intergalactic
medium. In two of the absorbing-galaxies presented here (Q0452$-$1640 and Q2352$-$0028), the \nii\ emission line is detected in the SINFONI data. These observations allow to build a  2D map of the \hii\ metallicity of the galaxies based on the N2-index as illustrated in Figure~\ref{f:Gradients}. The object towards Q0452$-$1640 has a rather uniform metallicity across the extent of the galaxy. Using a series of three distinct annuli with increasing radius, we derive a metallicity gradient of $-$0.11 dex/kpc across the object. The absorbing-galaxy towards Q2352$-$0028 is showing no sign of variation over the small region covered in this object. The dynamical range of the metallicity is small, leading to an estimated gradient of $-$0.07 dex/kpc. In addition, we note that the corresponding 2D metallicity map published by P\'eroux et al. (2011a) for the absorbing-galaxy towards Q1009$-$0026 also shows a weak gradient of $-$0.10 dex/kpc.

We refer the reader to P\'eroux et al. (2013b) for a full presentation of these results, and briefly mention here the findings which add to our understanding of the nature of gas flows. Cresci et al. (2010) have reported `inverse' (positive) gradients,
with the central, star forming regions having a lower metallicity than outer, less active ones, opposite to what is
seen in local galaxies. These authors conclude that the central gas has been diluted by the accretion of primordial gas, as expected from cold accretion models. Queyrel et al. (2012), however, advocate that inflows are preferably due to interactions rather than cold flows. In the objects we study, we find no strong evidence for gradients in the metallicity of the galaxy across the extent to which these ones are detected. Therefore, in all three cases, the gradients of metallicity with radius are rather flat and/or weakly negative and do not suggest the presence of infall of gas. We also note that these estimates are in line with the gradients reported in the local Universe and at z$\sim$1.2 in the MASSIV survey (Queyrel et al. 2009).

\begin{table*}
\begin{center}
\caption{{\bf Gas flows around the \nhi\ absorbers detected with SINFONI. } This table summarises the constraints from several indicators to test the various hypotheses on the direction of the flows around the galaxies: the orientation of the galaxy, its distance to the quasar line-of-sight, the orientation of the direction to the quasar line-of-sight with the galaxy's rotation axis, the maximum circular velocity, V$_{\rm max}$, the absorption profile velocity spread along the quasar line-of-sight, $\Delta v$ and a description of the absorption profile. In the cases of absorbing-galaxies towards Q1009$-$0026 and Q2222$-$0946, the evidences for the presence of an outflow traced in absorption are the strongest.}
\label{t:flows}
\begin{tabular}{ccccccccccccc}
\hline\hline
Quasar 		  &Galaxy 		&b	&Direction to quasar 	&V$_{\rm max}$	&$\Delta v$	&Absorption	&Conclusion\\
			&Orientation	&[kpc]	&line-of-sight aligned with	&[km/s]		&[km/s]		&Profile		&\\
\hline
Q0302$-$223	  &edge-on	&25	&minor axis	&11				&120		&doubled-peaked	&$\Rightarrow$co-rotating/outflow?\\	
Q0452$-$1640  &face-on?	&16	&major axis	&100			&230		&either-side of z$_{\rm gal}$ &$\Rightarrow$merger/outflow?\\
Q1009$-$0026  &edge-on	&39	&minor axis?	&250			&334		&asymmetrical   		&$\Rightarrow$outflow\\
Q2222$-$0946  &edge-on	&6	&n/a$^\dagger$&20				&200		&centred and complex  		&$\Rightarrow$outflow\\
Q2352$-$0028  &edge-on	&12	&major axis	&140			&220		&centred and complex 		&$\Rightarrow$co-rotating/outflow?\\
\hline
\hline
\end{tabular}			       			 	 
\end{center}			       			 	 
\vspace{0.2cm}
\begin{minipage}{180mm}
{\bf $^\dagger$:} in the case of Q2222$-$0946, the major axis is undefined because of the compact nature of the galaxy. \\
\end{minipage}
\end{table*}

\section{Discussion and Conclusion}

We have presented here new SINFONI data, at a resolution of $\sim$0.4-arcsec ($\sim$3 kpc at z$\sim$1), of five high-\nhi\ absorbers reported by P\'eroux et al. (2011a, 2012). Using these observations as well as our earlier lower resolution (0.8-arcsec) observations, we are able to derive the kinematics of these systems: two of the systems show indications of mergers. In addition, we have information on the morphology of the absorbing-galaxies: we find that the all four z$\sim$1 absorbing-galaxies are rotating, indicating the presence of discs. The system at z$\sim$2 is compact, showing no indication of rotation.

Furthermore, the data are used to measure the dynamical masses of the systems which are found to range from 10$^{9.8}$ to 10$^{10.9}$ M$_{\odot}$. The mass of gas, however, is found to be between 10$^{8.8}$ to 10$^{9.7}$ M$_{\odot}$. We note that whenever the halo masses have been derived, these are significantly larger than the gas masses (see Table~\ref{t:kine}). In addition, in the case of Q0302$-$223, where the absorbing-galaxy is associated with a rotating system, we note that the gas mass is 2.5 times smaller than the stellar mass which is unexpected for such galaxies. This can partly be explained by the fact that our calculation is solely accounting for gas associated with the visible region of the galaxy. Moreover, for the rotating galaxies, we are able to estimate the mass of the halo in which the absorbers reside assuming the systems are virialised. We find large values ranging from 10$^{11.8}$ to 10$^{12.8}$ M$_{\odot}$. These halo masses are an order of magnitude larger than the one derived by Pontzen et al. (2008) based on dedicated Smoothed-Particle Hydrodynamics (SPH) simulations. In fact, these authors predict that the major contributors to the population of DLAs are haloes of masses 10$^{9}$$<$M$_{\rm halo}$$<$10$^{11}$ $M_{\odot}$, with a peak at M$_{\rm halo}$=10$^{10}$ $M_{\odot}$. They predict 3 times fewer systems with M$_{\rm halo}$=10$^{11}$ $M_{\odot}$ than systems with M$_{\rm halo}$=10$^{10}$ $M_{\odot}$. It is possible that the 5 high-\nhi\ absorbers we have detected amongst 16 searched for are thus the high mass end of the DLA distribution. 

In addition, our measurements of the absorbing-galaxy cross-section and M$_{\rm halo}$ imply that the three systems reported here have halo masses 4 to 5 orders of magnitude larger or cross-sections 3 orders of magnitude smaller than required to lie on the relation predicted by Pontzen et al. (2008). In fact, these authors note that their estimate of the cross-sections are larger than in previous simulations (Gardner et al. 1997; Nagamine et al. 2004) and suggest that this might be due to their particular feedback implementation. Pontzen et al. (2008) also note that the DLAs with high-mass and small cross-section compose only few percent of the population in the simulations. While the systems observed here are relatively metal-rich and hence expected to have (or have had) large SFR, the detection rate of our sample indicates that theses are representative of more than just a few percent of the DLA population. Similarly, our estimates of the SFR in these systems correspond to halo masses 1.5 to 3 orders of magnitude larger than the one predicted by Pontzen et al. (2008) based on their predicted relation.

Finally, in two of the cases (Q0302$-$223 and Q2222$-$0946), additional HST imaging allows to determine the stellar mass of the absorbing-galaxies. These are derived to be 10$^{9.5}$ to 10$^{9.3}$ M$_{\odot}$, respectively (see Krogager et al. 2013). These measurements allow for a direct test of the mass-metallicity relation in quasar absorbers claimed by several authors (Ledoux et al. 2006; M\"oller et al. 2013). Using the relation of M\"oller et al. (2013) and our estimate of the emission metallicity (P\'eroux et al. 2013b), we derive a stellar mass of 10$^{9.2}$  M$_{\odot}$ for the absorbing-galaxy towards Q0302$-$223. Our measurements are therefore in line with the MMR reported by M\"oller et al. (2013). Krogager et al. (2013) have reached similar conclusions regarding the system towards Q2222$-$0946.

We further use several indicators to study the flow of gas around these absorbing-galaxies. Indeed, the observations presented here allow to combine various tests which, together, can put constrains on the directions of the flows around z$\sim$1 or 2 galaxies. Based on arguments on the star formation per unit area, we argue that all systems might produce winds. In all five cases, we measure EW(\mgii\ $\lambda$ 2796) $>$ 1\AA\ at the position of these absorbers and find that the saturated profiles extend both sides of the galaxies systemic redshifts, which are believed to be signatures of winds. Using a comparison of the emission and absorption kinematics, as well as inclination and orientation arguments, we find that two of the systems show signature of an outflow in the velocity profile. For the remaining three systems, it is difficult to reach definitive conclusions. In particular, there are two cases where the presence of two separate objects detected in HST and SINFONI imaging (Q0302$-$223 and Q0452$-$1640, respectively) complicate the interpretation. These results are summarised in Table~\ref{t:flows}. Finally, 2D abundance maps and measure of the metallicity gradients in three of the five systems do not indicate signatures expected from infall of fresh gas onto the galaxies. Overall, our data are therefore consistent with the gas seen in absorption being due to material co-rotating with the halo of their discs although some lines of evidence might support the presence of outflows traced in absorption. In the case of absorbing galaxies toward Q1009$-$0026 and Q2222$-$0946, we have the strongest evidence for the presence of outflows. This is also supported by a large value of star formation rate per unit area, $\Sigma_{\rm SFR}$ derived in the case of Q2222$-$0946.

\section*{Acknowledgements}
We would like to thank the Paranal and Garching staff at ESO for performing the observations in service mode and the instrument
team for making a reliable instrument. We are grateful to Joe Meiring, Debopam Som and Max Pettini for providing the high-resolution spectra of two of the targets and Johan Fynbo for communicating results on another object in advance of publication. 
CP thanks Thomas Ott for developing and distributing the QFitsView software. VPK acknowledges partial support from the U.S. National Science Foundation grants AST/0908890, AST/1108830 (PI: Kulkarni). This work has benefited from support of the 'Agence Nationale de la Recherche' with reference ANR-08-BLAN-0316-01.

\bsp

\label{lastpage}


\begin{thebibliography}{99}

\bibitem[]{} Cohn, H. \& York, D. G., 1977, ApJ, 216, 408

\bibitem[]{} Cowie, L. L. \& York, D. G., 1978, ApJ, 220, 129

\bibitem[]{} Cucciati, O., et al.,  2012, A\&A, 539A, 31

\bibitem[]{} Bouch\'e, N., Murphy, M. T.,  P\'eroux, C., Davies, R., Eisenhauer, F., F\"orster Schreiber, N. M. \& Tacconi, L., 2007a, ApJ, 669L, 5 

\bibitem[]{} Bouch\'e, N., et al., 2007b, ApJ, 671, 303

\bibitem[]{} Bouch\'e, N., Murphy, M. T.,  P\'eroux, C., Contini, T., Martin, C., F\"orster Schreiber, N. M., Lutz, D., Gillessen, S., Genzel, R. , Tacconi, L. , Davies, R. \& Eisenhauer, F., 2012, MNRAS, 419, 2

\bibitem[]{} Bouch\'e, N., Murphy, M. T., Kacprzak, G. G., P\'eroux, C., Contini, T., Martin, C. \& Dessauges-Zavadsky, M., 2013, Nature, 341, 40

\bibitem[]{} Bordoloi, R., et al., 2011, ApJ, 743, 10

\bibitem[]{} Cresci, G., Mannucci, F., Maiolino, R., Marconi, A., Gnerucci, A. \& Magrini, L., 2010, Nature, 467, 811

\bibitem[]{} Coil, A. L., Weiner, B. J., Holz, D. E., Cooper, M. C., Yan, R. \& Aird, J., 2011, ApJ, 743, 46

\bibitem[]{} Daddi, E., et al., 2010, ApJ, 713, 686

\bibitem[]{} Dav\'e, R. \& Oppenheimer, B., 2007, MNRAS, 374, 427

\bibitem[]{} Dekel, A. et al., 2009, Nature, 457, 451

\bibitem[]{} Epinat, B., et al., 2009, A\&A, 504, 789

\bibitem[]{} Erb, D. K., Shapley, A. E., Pettini, M., Steidel, C. C., Reddy, N. A. \& Adelberger, K. L., 2006, ApJ, 644, 813

\bibitem[]{} Finkelstein, S. L., 2009, ApJ, 700, 376

\bibitem[]{} F\"orster-Schreiber, N., M., et al., 2006, ApJ, 645, 1062

\bibitem[]{} Fynbo, J. P. U., Prochaska, J. X., Sommer-Larsen, J., Dessauges-Zavadsky, M. \& M\"oller, P., 2008, ApJ, 683, 321

\bibitem[]{} Fynbo, J. P. U., Laursen, P., Ledoux, C., Moller, P., Goldoni, P., Gullberg, B., Kaper, L., Maund,  J., Noterdaeme, P. , Ostlin, G., Strandet, M. L. , Toft, S., Vreeswijk, P. M. \& Zafar, T., 2010, MNRAS, 408, 2128

\bibitem[]{} Fynbo, J. P. U., Geier, S., Christensen, L., Gallazzi, A., Krogager, J.-K., KrŸhler, T., Ledoux, C., Maund, J., M\"oller, P., Noterdaeme, P., Rivera-Thorsen, T. \& Vestergaard, M., 2013 (arXiv1306.2940)

\bibitem[]{} Gardner, J. P., Katz, N., Weinberg, D. H. \& Hernquist, L., 1997, ApJ, 486, 42

\bibitem[]{} Gauthier, J.-R. \& Chen, H.-W., 2012, MNRAS, 424, 1952

\bibitem[]{} Genzel, R., et al., 2008, ApJ, 687, 59

\bibitem[]{} Genzel, R., et al., 2010, MNRAS, 407, 2091

\bibitem[]{} Genzel, R., et al.,  2011, ApJ, 733, 101

\bibitem[]{} Giavalisco, M., et al., 2011, ApJ, 743, 95

\bibitem[]{} Haehnelt, M. G., Steinmetz, M. \& Rauch, M., 1998, ApJ, 495, 647

\bibitem[]{} Heckman, T., 2002, Extragalactic Gas at Low Redshift, 254, 292

\bibitem[]{} Heckman, T., 2003, RMxAC, 17, 47

\bibitem[]{} Kacprzak, G. G., Churchill, C. W., Ceverino, D., Steidel, C. C., Klypin, A. \& Murphy, M. T., 2010, ApJ, 711, 533

\bibitem[]{} Kacprzak, G. G., Churchill, C. W., Evans, J. L., Murphy, M. T. \& Steidel, C. C.,  2011b, MNRAS, 416, 3118

\bibitem[]{} Kacprzak, G. G., Churchill, C. W., Barton, E. J. \&Cooke, J., 2011a, ApJ, 733, 105

\bibitem[]{} Kacprzak, G. G., Churchill, C. W., Steidel, C. C., Spitler, L. R. \& Holtzman, J. A., 2012b, MNRAS, 427, 3029

\bibitem[]{} Kacprzak, G. G., Churchill, C. W., Nielsen, N. M., 2012a, MNRAS, 427, 2711

\bibitem[]{} Kennicutt, R. C., Jr., 1998, ApJ, 498, 541

\bibitem[]{} Kobayashi, C., Springel, V. \& White, S. D. M.,  2007, MNRAS, 376, 1465

\bibitem[]{} Kornei, K. A., et al., 2012, ApJ, 758, 135

\bibitem[]{} Krogager, J-H, Fynbo, J., Ledoux, C., Christensen, L., Gallazzi, A., Laursen, P., M\"oller, P., Noterdaeme, P., Peroux, C., Pettini, M., Vestergaard, M., 2013 (arXiv1304.4231)

\bibitem[]{} Law, D. R., Steidel, C. C., Shapley, A. E., Nagy, S. R., Reddy, N. A. \& Erb, D. K., 2012, ApJ, 759, 29

\bibitem[]{} Le Brun, V., Bergeron, J., Boisse, P., Deharveng, J. M., 1997, A\&A, 321, 733

\bibitem[]{} Ledoux, C., Petitjean, P., Fynbo, J. P. U., M\"oller, P. \& Srianand, R., 2006, A\&A, 457, 71

\bibitem[]{} Lequeux, J., Peimbert, M., Rayo, J. F., Serrano, A. \& Torres-Peimbert, S., 1979, A\&A, 80, 155

\bibitem[]{} Maller, A. H., Prochaska, J. X., Somerville, R. S. \& Primack, J. R., 2001, MNRAS, 326, 1475

\bibitem[]{} Martin, C. L. \& Bouch\'e, N., 2009, ApJ, 703, 1394

\bibitem[]{} Martin, C. L., Shapley, A. E., Coil, A. L., Kornei, K. A., Bundy, K., Weiner, B. J., Noeske, K. G. \& Schiminovich, D., 2012, ApJ, 760, 127

\bibitem[]{} Martin, C. L., Shapley, A. E., Coil, A. L., Kornei, K. A., Murray, N. \& Pancoast, A., 2013, ApJ, 770, 41

\bibitem[]{} Meiring, J. D., Lauroesch, J. T., Kulkarni, V. P., P\'eroux, C., Khare, P., York, D. G. \& Crotts, A. P. S., 2007, MNRAS, 376, 557 
 
\bibitem[]{} Meiring, J. D., Kulkarni, V. P., Lauroesch, J. T., P\'eroux, C., Khare, P. \& York, D. G., 2009, MNRAS, 393, 1513

\bibitem[]{} Mo, H. J. \& White, S. D. M.,  2002, MNRAS, 336, 112 

\bibitem[]{} M\"oller, P., Fynbo, J. P. U., Ledoux, C. \& Nilsson, K. K., 2013, MNRAS, 430, 2680

\bibitem[]{} Murray, N., M\'enard, B. \&Thompson, T. A., 2011, ApJ, 735, 66

\bibitem[]{} Nagamine, K., Springel, V. \& Hernquist, L., 2004, MNRAS, 348, 421

\bibitem[]{} Nelson, E. J., van Dokkum, P. G., Brammer, G., F\"orster Schreiber, N., Franx, M., Fumagalli, M., Patel, S., Rix, H-W., Skelton, R. E., Bezanson, R., Da Cunha, E., Kriek, M., Labbe, I., Lundgren, B., Quadri, R. \& Schmidt, K. B., 2012, ApJ, 747L, 28

\bibitem[]{} Newman, S. F., et al., 2012, ApJ, 761, 43

\bibitem[]{} Noterdaeme, P., Petitjean, P., Ledoux, C. \& Srianand, R., 2009, A\&A, 505, 1087

\bibitem[]{} Noterdaeme, P., et al., 2012, A\&A, 547L, 1

\bibitem[]{} P\'eroux, C., Dessauges-Zavadsky, M., D'Odorico, S., Kim,T. S., \& McMahon, R., 2003, MNRAS, 345, 480.

\bibitem[]{} P\'eroux, C., Meiring, J., Kulkarni, V. P., Khare, P., Lauroesch, J. T., Vladilo, G. \& York, D. G., 2008, MNRAS, 386, 2209

\bibitem[]{} P\'eroux, C., Bouch\'e, N., Kulkarni, V. P., York, D. G. \& Vladilo, G., 2011a, MNRAS, 410, 2237

\bibitem[]{} P\'eroux, C., Bouch\'e, N., Kulkarni, V. P., York, D. G. \& Vladilo, G., 2011b, MNRAS, 410, 2251

\bibitem[]{} P\'eroux, C., Bouch\'e, N., Kulkarni, V. P., York, D. G. \& Vladilo, G., 2012, MNRAS, 410, 2251

\bibitem[]{} P\'eroux, C., Kulkarni, V. P. \& York, D. G.,  2013b, MNRAS, submitted

\bibitem[]{} Petitjean, P. \& Bergeron, J., 1990, A\&A, 231, 209

\bibitem[]{} Pettini, M., Ellison, S. L., Steidel, C. C., Shapley, A. E. \& Bowen, D. V., 2000, ApJ, 532, 65

\bibitem[]{} Pettini, M., Madau, P., Bolte, M., Prochaska, J. X., Ellison, S. L. \& Fan, X., 2003, ApJ, 594, 695

\bibitem[]{} Pontzen, A., Governato, F., Pettini, M., Booth, C. M., Stinson, G., Wadsley, J., Brooks, A., Quinn, T. \& Haehnelt, M., 2008, MNRAS, 390, 1349

\bibitem[]{} Prochaska, J. X., Herbert-Fort, S. \& Wolfe, A. M., 2005, ApJ, 635, 123

\bibitem[]{} Putman, M.E., et al. 2009, "How do galaxies accrete gas and form stars?", The Astronomy \& Astrophysics Decadal Survey 2010 

\bibitem[]{} Queyrel, J., et al., 2009, A\&A, 506, 681

\bibitem[]{} Queyrel, J., et al., 2012, A\&A, 539A, 93

\bibitem[]{} Rubin, K. H. R., Weiner, B. J., Koo, D. C., Martin, C. L., Prochaska, J. X., Coil, A. L. \& Newman, J. A., 2010, ApJ, 719, 1503

\bibitem[]{} Ryan-Weber, E. V., Pettini, M., Madau, P. \& Zych, B. J., 2009, MNRAS, 395, 1476

\bibitem[]{} Sato, T., Martin, C. L., Noeske, K. G., Koo, D. C. \& Lotz, J. M., 2009, ApJ, 696, 214

\bibitem[]{} Savaglio, S., Glazebrook, K., Le Borgne, D., Juneau, S., Abraham, R. G., Chen, H.-W., Crampton, D., McCarthy, P. J., Carlberg, R. G., Marzke, R. O., Roth, K., J\"orgensen, I. \& Murowinski, R.,  2005, ApJ, 635, 260 

\bibitem[]{} Shen, S., Madau, P., Guedes, J., Mayer, L., Prochaska, J. X. \& Wadsley, J., 2013, ApJ, 765, 89

\bibitem[]{} Steidel, C. C., et al., Kollmeier, J. A., Shapley, A. E., Churchill, C. W., Dickinson, M. \& Pettini, M., 2002, ApJ, 570, 526

\bibitem[]{} Steidel, C. C., et al., 2010, ApJ, 717, 289

\bibitem[]{} Steidel, C. C., Bogosavljevi\'c, M., Shapley, A. E., Kollmeier, J. A., Reddy, N. A., Erb, D. K. \& Pettini, M., 2011, ApJ, 736, 160

\bibitem[]{} Stewart, K. R., Kaufmann, T., Bullock, J. S., Barton, E. J., Maller, A. H., Diemand, J. \& Wadsley, J., 2011, ApJ, 738, 39

\bibitem[]{} Stinson, G., S., et al., 2012, MNRAS, 425, 1270

\bibitem[]{} Tremonti, C. A., et al.,  2004, ApJ, 613, 898

\bibitem[]{} Tremonti, C. A., Moustakas, J. \& Diamond-Stanic, A. M., 2007, ApJ, 663L, 77

\bibitem[]{} van Dokkum, P. G., 2001, PASP, 113, 1420

\bibitem[]{} Weiner, B. J. et al., 2009, ApJ, 692, 187

\bibitem[]{} White, S. D. M. \& Rees, M. J., 1978, MNRAS, 183, 341

\bibitem[]{} Wright, S. A., Larkin, J. E., Law, D. R., Steidel, C. C., Shapley, A. E. \& Erb, D. K., 2009, ApJ, 699, 421 

\bibitem[]{} York, D. et al., in prep.

\bibitem[]{} Zafar, T., P\'eroux, C., Popping, A., Deharven, J.-M., Milliard, B. \& Frank, S., 2013, MNRAS, submitted

\bibitem[]{} Zwaan, M., Walter, F., Ryan-Weber, E., Brinks, E., de Blok, W. J. G. \& Kennicutt, R. C., Jr., 2008, AJ, 136, 2886

\end{thebibliography}
\end{document}